\documentclass[12pt]{article}
\pdfoutput=1

\usepackage{graphicx,array}
\usepackage{color}
\usepackage{latexsym}
\usepackage{amsthm}
\usepackage{amsmath}
\usepackage{amssymb}
\usepackage[numbers,sort&compress]{natbib}
\usepackage{bm}
\usepackage{comment} 
\usepackage{slashed}
\usepackage{mathrsfs}
\usepackage{latexsym}
\usepackage[font={small,it}]{caption}
\usepackage{hyperref} 
\hypersetup{
    colorlinks=true,       
    linkcolor=red,          
    citecolor=blue,        
    filecolor=magenta,      
    urlcolor=blue           
}
\usepackage[all]{hypcap} 

\usepackage{myterms}
\newcommand{\PQ}{\text{\sc pq}}
\newcommand{\DW}{{\rm DW}}

\newcommand{\RH}{\text{\sc rh}}
\newcommand{\be}{\begin{equation}}
\newcommand{\ee}{\end{equation}}

\title{Cosmological Aspects of the Clockwork Axion}

\author{\large Andrew J. Long}
\date{\small \it 
Kavli Institute for Cosmological Physics, University of Chicago, Chicago, Illinois 60637, USA 
}

\begin{document}

\maketitle

\setlength{\parskip}{0.2ex}

\begin{abstract}
The clockwork axion refers to a family of aligned multi-axion models that lead to an exponential hierarchy between the scale of Peccei-Quinn symmetry breaking and the scale of the axion decay constant.  
The clockworking can bring the Peccei-Quinn-scale particles to within reach of collider experiments.
In this work we are interested in whether cosmological observations impose any new constraints on the clockwork axion.  
If the universe reheats above the scale of Peccei-Quinn breaking, then the ensuing cosmological phase transition produces a network of topological defects, which have a qualitatively different behavior from the string-wall network in the usual axion models.  
We estimate the relic abundances of axion dark matter and dark radiation that arise from the emission of axions by the defect network, and we infer a constraint on the scale of Peccei-Quinn breaking and the mass spectrum.  
We find that the defect contribution to the axion {\it dark matter} relic abundance is generally negligible.  
However, the defect production of relativistic axion {\it dark radiation} becomes significant if the scale of Peccei-Quinn symmetry breaking is larger than $100 \, {\rm TeV}$, and measurements of $\Delta N_{\rm eff}$ provide a new probe of this class of models.  
\end{abstract}

\newpage
\section{Introduction}
\label{sec:Introduction}

The axion \cite{Peccei:1977hh, Peccei:1977ur, Weinberg:1977ma, Wilczek:1977pj,Kim:1979if,Shifman:1979if,Dine:1981rt,Zhitnitsky:1980tq} is an elegant solution to the strong CP problem that also naturally provides a candidate for the dark matter.  
(See \rref{Kuster:2008zz} for a general review.)
The axion solution introduces new particles and interactions at a scale, $f_\PQ$, set by the spontaneous breaking of the global $\U{1}$ Peccei-Quinn (PQ) symmetry, and the axion is the corresponding pseudo-Goldstone boson.  
Generally $f_\PQ$ is well above the weak scale, and upon integrating out the PQ-scale particles, the axion acquires dimension-five interactions with the gluons (and possibly other Standard Model particles as well) with a coupling $g_{agg} \equiv \alpha_s / (2 \pi f_g)$, which defines the parameter $f_g$.  
Due to instanton effects in quantum chromodynamics (QCD), this interaction induces a potential for the axion, which lifts its mass to \cite{veneziano,diCortona:2015ldu}
\be\label{eq:ma}
m_a \simeq 5.6 \, \mu{\rm eV} \left( \frac{f_g}{10^{12} \GeV} \right)^{-1}
\per 
\ee
In standard axion models one usually finds $f_g = f_\PQ / N_\DW$, where the QCD domain wall number, $N_\DW$, is typically an $O(1)$ integer that grow linearly with the number of PQ-scale colored fermions.  
Consequently, the most well-studied axion models predict the scale of the axion-gluon coupling and the scale of PQ breaking to be comparable, $f_g \sim f_\PQ$.  

In this work, we study the cosmological implications of the following relation: 
\be\label{separation}
f_\PQ = z f_g \quad \mathrm{with}\quad  z\ll 1 \quad \text{and} \quad N_\DW=1 
\per 
\ee
In general aligned axion models with multiple pseudoscalar fields \cite{Kim:2004rp} allow one to break the canonical relation between $f_\PQ$ and $f_g$.   
The clockwork mechanism \cite{Choi:2014rja,Choi:2015fiu,Kaplan:2015fuy} provides a concrete benchmark model in which \eref{separation} can be realized with an exponential hierarchy.  
The mechanism, which will be reviewed in \sref{sec:Clockwork}, introduces $N+1$ complex scalar fields with global $\U{1}$ symmetries that are spontaneously broken at a scale $f_\PQ$ generating $N+1$ Goldstone bosons.  
Additionally, a specific scalar potential explicitly breaks $N$ of the symmetries lifting all of the flat directions save one, which corresponds to an axion with axion-gluon coupling strength set by $f_g \sim q^N f_\PQ$ for integer $q$.  
The model with $q = 3$ is renormalizable and well-defined \cite{Kaplan:2015fuy}, and we will use it for all the studies in this paper.

The regime $f_\PQ \ll f_g$ is particularly interesting for both particle phenomenology and cosmology.  
Whereas $f_g$ must remain larger than roughly $10^8 \GeV$ to satisfy astrophysical constraints on axion-gluon interactions (for a review see \rref{Marsh:2015xka}), the scale of PQ breaking could be as low as $f_\PQ \sim 1 \TeV$.  
This puts the new PQ-scale particles within reach of collider experiments \cite{Farina:2016tgd}.  
If the primordial plasma reaches temperatures above $f_\PQ$, which is quite reasonable for compelling models of inflation and reheating, then PQ-symmetry breaking is accomplished through a cosmological phase transition.  
It is well known from standard QCD axion models that the PQ-breaking phase transition produces a network of topological defects \cite{VilenkinShellard:1994}, namely global strings and domain walls, which efficiently radiate the Goldstone axion \cite{Vilenkin:1982ks,Vachaspati:1984yi,Davis:1986xc,Vilenkin:1986ku,Harari:1987ht} and contribute to the axion relic abundance.  
Previous studies of these defect networks have generally assumed $f_\PQ \sim f_g$, but this relation is badly broken for the clockwork axion, and moreover, the presence of $N+1$ complex scalar fields participating in the phase transition gives the defect network a much richer structure than what is found in the usual QCD axion cosmology.  

In this article we are interested in the structure of the defect network and its contribution to the axion dark matter and dark radiation relic abundances.  
For the usual QCD axion, the collapse of the defect network at the QCD epoch ($t = t_\QCD$) is one of the dominant contributions to the axion dark matter relic abundance \cite{Marsh:2015xka}, which corresponds to an energy density $\rho_a(t_\QCD) \sim \varepsilon \, \sigma H_\QCD$ where $\varepsilon$ measures how efficiently axions are emitted from the domain walls, $\sigma \sim m_a f_g^2$ is the surface tension of the axion-field domain walls, and $H_\QCD = H(t_\QCD)$ is the Hubble parameter.  
Since the domain walls are composed of the axion field, the emission of axion particles is very efficient and typically $\varepsilon \sim 1$ \cite{Vilenkin:1982ks,Vachaspati:1984yi,Davis:1986xc,Vilenkin:1986ku,Harari:1987ht}.  
In the clockwork axion model we will see that the domain walls are built from the additional pseudoscalar (gear) fields with mass $m_G$, and the wall tension is $\sigma \sim m_G f_\PQ^2$.  
Therefore it is reasonable to guess $\rho_a(t_\QCD) \sim \varepsilon^\prime m_G f_\PQ^2 H_\QCD$, which corresponds to a relic abundance of axion dark matter given by 
\begin{align}\label{eq:Oah2_naive}
	\Omega_a h^2 \sim \varepsilon^\prime \left( \frac{m_G}{100 \GeV} \right) \left( \frac{f_\PQ}{10 \TeV} \right)^2
	\qquad \text{(naive estimate)}
	\per
\end{align}
This estimate suggests that the observed dark matter relic abundance ($\Omega_\DM h^2 \simeq 0.12$) can be explained if the clockworking lowers the PQ scale to within reach of collider experiments.  
Moreover, \eref{eq:Oah2_naive} suggests a strong upper bound on $m_G$ and $f_\PQ$ in order to avoid producing too much axion dark matter.  
However, it is not immediately obvious how large is the factor $\varepsilon^\prime$, which represents the efficiency with which axions are emitted from the domain walls, which are now composed of the gear fields rather than the axion field itself.  

To confirm or refute the naive estimate, we have performed a more careful calculation of the axion emission from the clockwork defect network.  
Since the pseudoscalar (gear) fields that form the domain walls only have a weak coupling to the axion, it is found that axion emission in the clockwork model is generally much less efficient than for the usual QCD axion where the axion field itself composes the walls; that is, \eref{eq:epsilon_prime} reveals that $\varepsilon^\prime \ll \varepsilon \sim 1$ for Hubble-scale domain walls at the QCD epoch.  
The result is a suppression of the axion dark matter relic abundance arising from the defect network.  
(The relic abundance from the misalignment mechanism depends on the axion decay constant $f_g$ directly, and it is unaffected by the clockworking.)  
On the other hand, we also find that axion emission from small-scale structure on the domain walls is still efficient, and this can give rise to a population of relativistic axions that contribute to the dark radiation.  

The remainder of this article is organized in the following way.  
We briefly review the clockwork axion in \sref{sec:Clockwork} with an emphasis on understanding the topology of the field space.  
In \sref{sec:Cosmology} we focus on the cosmology of the topological defect network and its contribution to the axion dark matter and dark radiation relic abundances.  
We summarize our results in \sref{sec:Conclusion} and briefly discuss some more general cosmological implications.  


\section{The Clockwork Axion}
\label{sec:Clockwork}

The clockwork axion is a class of multi-axion models \cite{Kim:2004rp} with an exponentially large hierarchy between the PQ-breaking scale, $f_\PQ$, and the axion decay constant, $f_a$.  
Different implementations were explored in the original literature \cite{Choi:2014rja,Choi:2015fiu,Kaplan:2015fuy} and subsequent work \cite{Giudice:2016yja,Giudice:2017fmj,Giudice:2017suc,Craig:2017cda,Craig:2017ppp,Teresi:2018eai}.  
In this section we briefly review the model that was proposed in \rref{Kaplan:2015fuy}.  

The model consists in a family of $N+1$ complex scalar fields that are denoted by $\phi_n(x)$ with $n \in \{ 0, 1, \cdots, N \}$.  
The properties and interactions of these fields are determined by the Lagrangian\footnote{\eref{eq:Lagrangian} has an approximate (flavor) permutation symmetry acting on the $\phi_n$. This symmetry is a consequence of not allowing for generic $\lambda_n$, $\epsilon_n$, $f_n$, and in general need not to be present. However, it simplifies the model and the description of the defects dynamics in the early universe.  The symmetry is broken by the $\epsilon$-term and by the coupling to PQ matter fields. When the dynamics is perturbative in $\epsilon$ and $y$, we will exploit the flavor symmetry to understand the scaling of some observables with the parameter of the models. The discussion of the effects of the breaking of this symmetry can be found in \cite{Craig:2017cda}.}
\begin{align}\label{eq:Lagrangian}
	\mathscr{L} = \sum_{n=0}^N \Bigl[ \bigl| \partial_\mu \phi_n \bigr|^2 - \lambda \bigl( | \phi_n |^2 - f_\PQ^2 / 2 \bigr)^2 \Bigr] + \epsilon \sum_{n=0}^{N-1} \Bigl[ \phi_n^\ast \phi_{n+1}^3 + \hc \Bigr] 
	\per 
\end{align}
The four model parameters are the dimensionless couplings $\lambda$ and $\epsilon$, the energy scale $f_\PQ$, and the integer $N$.
The first term in the Lagrangian \pref{eq:Lagrangian} is invariant under the symmetry  
\begin{align}\label{eq:U1_N+1}
	\U{1}^{N+1} \ : & \quad \phi_n \to {\rm exp}\bigl[i \theta_n\bigr] \phi_n 
\end{align}
with $\theta_n \in [0,2\pi)$ for $n \in \{0,1,\cdots,N\}$.  
The second term in \eref{eq:Lagrangian} explicitly breaks this symmetry down to the subgroup
\begin{align}\label{eq:U1_PQ}
	\U{1}_\PQ \ : & \quad \phi_n \to {\rm exp}\bigl[ i q^{N-n} \theta \bigr] \phi_n 
\end{align}
with $\theta \in [0,2\pi)$ and $q\equiv3$.  
We see that field $\phi_n$ has charge $q^{N-n}$ under the $\U{1}_\PQ$ symmetry.  
Notice also that the $\U{1}_\PQ$ symmetry has a family of discrete subgroups, 
\begin{align}\label{eq:Zq}
	\Zbb_{q^{N-k}}:\,\quad \phi_n \to {\rm exp}\bigl[ i 2\pi q^{k-n}  \bigr] \phi_n 
	\com 
\end{align}
which correspond to taking $\theta = 2\pi q^{k-N}$ with $k \in \{ 0, 1, \cdots, N-1 \}$ in \eref{eq:U1_PQ}.  

The scalar potential in \eref{eq:Lagrangian} leads to spontaneous symmetry breaking.  
The $N+1$ complex scalar fields acquire nonzero vacuum expectation values $\langle \phi_n \rangle = (f_\PQ / \sqrt{2}) \bigl( 1 + O(\epsilon) \bigr)$ where the $\epsilon$-dependent shift is not independent of $n$.  
Consequently the $\U{1}^{N+1}$ symmetry is spontaneously broken without leaving any unbroken subgroup.  
Since the $\epsilon$-dependent term in \eref{eq:Lagrangian} explicitly breaks $\U{1}^{N+1}$ down to $\U{1}_\PQ$, the spectrum contains $N$ massive pseudo-Goldstone bosons and $1$ massless Goldstone boson.  

\begin{figure}[t]
\begin{center}
\includegraphics[width=0.95\textwidth]{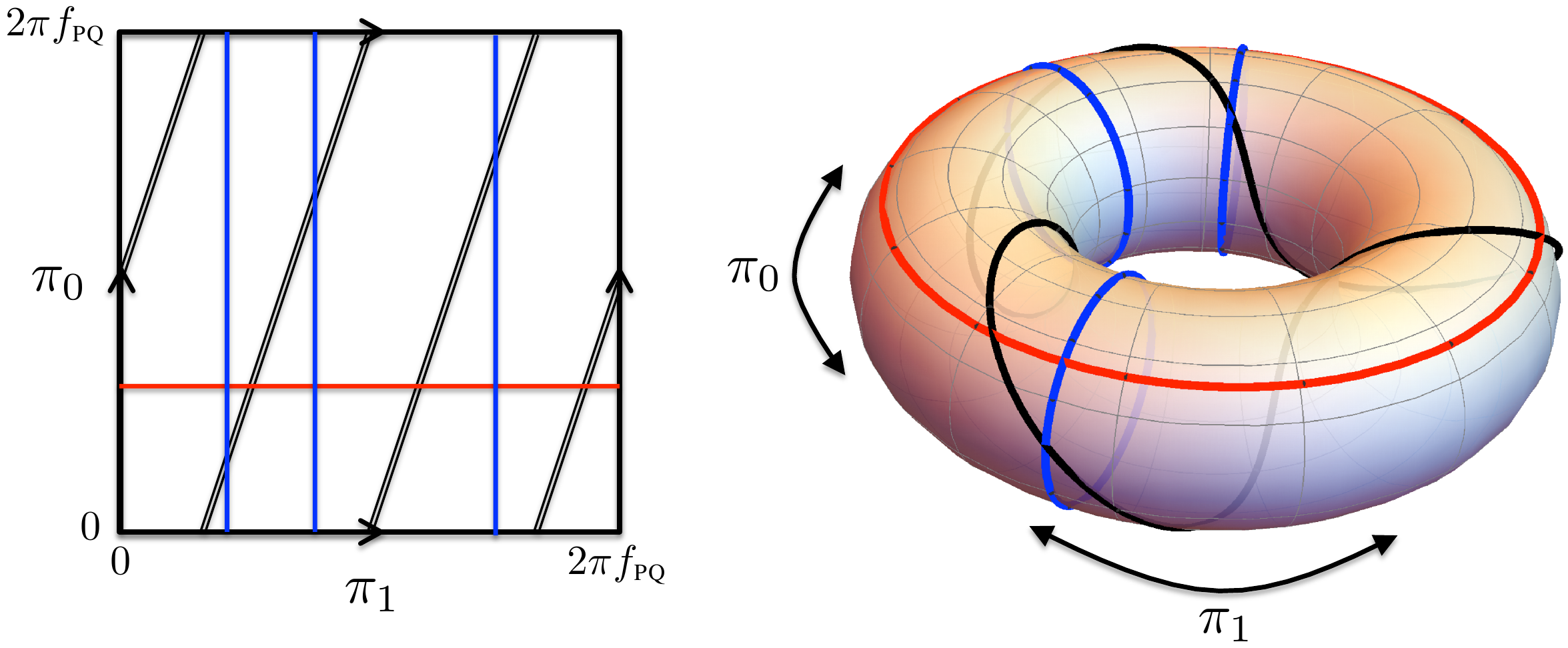} 
\caption{\label{fig:config_space}
For the model with $N+1 = 2$ complex scalar fields the pseudoscalar field space $(\pi_0, \pi_1)$ is topologically equivalent to a 2-torus.  
For $\epsilon = 0$ and QCD instanton effects neglected, there is no potential for $(\pi_0, \pi_1)$; taking $\epsilon \neq 0$ induces a potential, leaving a single flat direction, which is represented by the black line; and at the QCD epoch, instanton effects lift also this flat direction.  
Along the flat direction (black line) the axion field covers $a \in [0, 2\pi f_a)$ \pref{eq:clockworking} while the field $\pi_n$ completes $q^{N-n}$ cycles \pref{eq:pi_n} corresponding to its PQ charge \pref{eq:U1_PQ}.  
The three blue lines indicate three $\pi_0$-strings, each connected to one domain wall; the red line indicates a single $\pi_1$-string connected to three domain walls; and the black line indicates an $a$-string.  
See discussion in \sref{sec:strings}.  
}
\end{center}
\end{figure}


In order to understand the vacuum structure of this theory, it is useful to parametrize the complex scalar fields $\phi_n$ in terms of real scalar fields $\rho_n$ and the pseudoscalar fields $\pi_n$ by writing 
\begin{align}\label{eq:phi_n}
	\phi_n = \frac{1}{\sqrt{2}} \bigl( f_\PQ + \rho_n \bigr) {\rm exp}\bigl[ i \pi_n / f_\PQ \bigr] 
	\per 
\end{align}
The full configuration space is covered when $\rho_n(x) \in [-f_\PQ, \infty)$ and $\pi_n(x) \in [0, 2\pi f_\PQ)$.  
Note that the Lagrangian is left invariant under the discrete gauge symmetries that take $\pi_n \to \pi_n + 2 \pi f_\PQ$, which correspond to the trivial transformations $\phi_n \to e^{2\pi i} \phi_n = \phi_n$.  
After modding out by this discrete gauge symmetry, the configuration space for the $\pi_n$ becomes compact.  
For instance, if $N=0$ then the configuration space of $\pi_0$ is topologically equivalent to a circle; if $N=1$ then the configuration space of $(\pi_0,\pi_1)$ is a torus; and so on.  
The $N=1$ case is illustrated in \fref{fig:config_space}.  


In the vacuum where $\langle \phi_n \rangle = f_\PQ / \sqrt{2}$ we identify the two mass parameters, 
\begin{align}\label{eq:mrho_mG}
	m_\rho \equiv \sqrt{2\lambda} \, f_\PQ
	\qquad \text{and} \qquad 
	m_G \equiv \sqrt{\epsilon/2} \, f_\PQ 
	\com 
\end{align}
which set the scales of the real scalar and pseudoscalar mass spectra.  
The $N+1$ real scalar fields $\rho_n$ have masses on the order of $m_\rho$ with a splitting on the order of $m_G$.  
We consider more carefully the spectrum of the $N+1$ pesudoscalar fields $\pi_n$, since the massless mode will be identified with the QCD axion.  
Let us denote the mass eigenstate pseudoscalar fields by $A_i$; we identify the massless axion as $a \equiv A_0$ and the $N$ massive gears as $G_i \equiv A_i$ for $i \in \{ 1, 2, \cdots, N \}$.  
A linear transformation relates the original field basis $\pi_n$ and the mass-diagonalized field basis $A_i$: 
\begin{align}\label{eq:pi_n}
	\pi_n 
	= \sum_{i=0}^{N} O_{ni} \, A_i 
	= O_{n0} \, a + \sum_{i=1}^{N} O_{ni} \, G_i 
	\com 
\end{align}
where one should remember that $\pi_n$ are evaluated modulo the discrete gauge symmetry $\pi_n \to \pi_n + 2 \pi f_\PQ$.  
The elements of the orthogonal matrix $O$ are given by \cite{Giudice:2016yja}
\begin{align}\label{eq:On0_Oni}
	O_{n0} = \frac{C}{q^n} 
	\qquad \text{and} \qquad 
	O_{ni} \sim 1 / \sqrt{N+1}
\end{align}
where $q = 3$ and $C \approx \sqrt{8/9} \simeq 0.94$.  
The pseudoscalar mass spectrum is given by 
\begin{align}\label{eq:mGi}
	m_{G_i} = \Bigl( q^2 + 1 - 2 q \, \cos \frac{i \pi}{N+1} \Bigr)^{1/2} m_G 
	\qquad \text{and} \qquad 
	m_a =0
\end{align}
where $G_i$ for $i \in \{1, 2, \cdots, N \}$ are the $N$ pseudoscalar gear fields, and $a$ is the massless axion.  

The specific wave-function for the massless mode ensures a very large field range for the axion. Indeed, since we must allow $\pi_n(x) \in [0, 2\pi f_\PQ)$ in order to span the full configuration space,
the property $O_{n0} \ll 1$ for $n = N \gg 1$ requires the axion field range be exponentially large.  
This implies that the axion field has a range 
\begin{align}\label{eq:clockworking}
	a \in [0, 2\pi f_a)
	\qquad \text{where} \qquad 
	f_a \equiv \frac{f_\PQ}{O_{N0}} = \frac{q^N f_\PQ}{C} 
	\per 
\end{align}
In other words, the transformation $a \to a + 2 \pi f_a$ is a discrete gauge symmetry, {\it e.g.} a field configuration with $a \geq 2\pi f_a$ is redundant with a configuration that has $0 \leq a < 2\pi f_a$.  
(Similar considerations give the allowed field range for the gears, $G_i$.)  For the case $N=1$ this calculation can be visualized concretely as in \fref{fig:config_space}.
Due to the clockworking \pref{eq:clockworking}, we can have $f_a \sim 10^{9} - 10^{11} \GeV$ in the classic axion mass window with $f_\PQ \sim 10 \TeV$ and $N \sim 10-15$, which potentially puts the new PQ-scale particles within reach of collider experiments.  

In order to interpret the Goldstone boson $a$ as the QCD axion, we must let the $\phi_n$ couple to colored fermions.  
Let $\psi_r$ and $\psi_{r^c}$ represent a pair of vector-like colored fermions in representations $r$ and $r^c$ of $\SU{3}_c$.  
We introduce a Yukawa coupling between these fermions and the scalar field at site-$l$, which is written as $y \phi_l \psi_r \psi_{r^c}$.  
As observed in \cite{Farina:2016tgd}, a phenomenologically ``safe'' choice corresponds to localizing the QCD fermions on the last site of the chain, which corresponds to $l=N$.  
Any other choice would lead to an exponentially large domain wall number $N_\DW$ (see below).  
Therefore, for $l=N$, integrating out the heavy colored fermions, $\psi_r$ and $\psi_{r^c}$, induces the color anomaly, which is 
\be\label{eq:piN_GG}
	\mathscr{L} \supset \frac{\alpha_s}{8\pi} \, \bigl( 2 d_\psi T(\psi) \bigr) \, \frac{\pi_N}{f_\PQ} \, G_{\mu\nu}^a \tilde G^{\mu\nu, a} 
\ee
where $T(\psi)$ is the index of the color representation ($T(\mathbf{3})=1/2$), and $d_\psi$ the dimension of the $\SU{2}_L$ representation ($d_\psi = 1$ for singlet).  
Writing $\pi_N$ in terms of the axion $a$ using Eqs.~(\ref{eq:pi_n}),~(\ref{eq:On0_Oni}),~and~(\ref{eq:clockworking}), and setting the gear fields to zero, the color anomaly becomes 
\be\label{eq:QCDNdw}
	\mathscr{L} \supset \frac{\alpha_s}{8\pi} \, \frac{a}{f_g} \, G_{\mu\nu}^a \tilde G^{\mu\nu, a}
	\qquad
	\text{with} 
	\qquad 
	f_g \equiv f_a \, / \, \underbrace{2 d_\psi T(\psi)}_{N_\DW} 
	\per 
\ee
The mismatch between the period of the discrete gauge symmetry ($2\pi f_a$) and the period of the effective QCD theta parameter ($2\pi f_g$) is the domain wall number, $N_\DW$.  
A reasonable choice is to consider $\psi_r$ in the same representation as the $u_R$ or $d_R$ quarks, which gives $N_\DW=1$ and avoids color stable relics with mass $m_\psi = y f_\PQ$, which are a potential disaster if PQ breaks after inflation. With this setup we can compute the axion mass; it is given by \eref{eq:ma} where $f_g=f_a$.


\subsection{Comparison with other multi-axion models}
\label{sec:comparision}

In this work, we focus on the model described by \eref{eq:Lagrangian}, since it is perhaps the simplest implementation of the clockwork axion.  
However, there exist compelling variations on the minimal model that do not require the introduction of $N \approx 10$ fundamental scalar fields.  
The low energy particle phenomenology is not appreciably changed from the minimal model to these variations, but the cosmology can differ.  
We briefly discuss generalizations and their implications in this section.  

Several alternative UV extensions of the clockwork axion were recently suggested by the authors of \rref{Agrawal:2017cmd}.  
The common feature of these variations is that the explicit symmetry-breaking term is allowed to arise dynamically.  
In effect, the parameter $\epsilon$ in \eref{eq:Lagrangian} becomes time dependent.  
A first implementation supposes that the symmetry-breaking ($\epsilon$) term arises from strong dynamics.  
In this scenario, $\epsilon$ is effectively zero at high energy scales, but it takes on a fixed, nonzero value below a certain energy scale corresponding to the confinement of the new strong force.  
A second implementation supposes that the symmetry-breaking term arises from a Froggat-Nielsen-type mechanism.  
The $\epsilon$ term is replaced by a non-renormalizable operator, such as $\sum_n \phi_n^\ast ( \Sigma_{n,n+1} / \Lambda )\phi_{n+1}^3$ where $\Lambda$ is the cutoff of the effective theory, and $\Sigma_{n,n+1}$ transforms as a bi-fundamental.  
After $\Sigma$ gets a vacuum expectation value, the symmetry-breaking term in \eref{eq:Lagrangian} is generated with $\epsilon \sim \langle \Sigma \rangle / \Lambda$.  
This hypothesis naturally accommodates small values, $\epsilon \ll 1$, since it is related to the ratio of hierarchical energy scales.  

Provided that the new particles and interactions are not within reach of collider experiments, then the predictions for particle physics observables are not significantly changed from the predictions of the effective theory in \eref{eq:Lagrangian}.  
However, the cosmology can be somewhat different, because $\epsilon$ is effectively time dependent.  
If PQ-symmetry breaking occurs while $\epsilon = 0$ then the network of topological defects is consists of cosmic strings, without any domain walls.  
This is simpler than the scenario we describe next in \sref{sec:Cosmology} where the strings are bound to walls during the prior from PQ breaking until the QCD epoch . 
Provided that $\epsilon$ becomes nonzero before the QCD epoch, and assuming that $m_G > H$ at this time, the strings become bound by domain walls.  
Then the predictions for cosmological observables, like the axion relic abundance, are not significantly changed from the scenario we describe in \sref{sec:Cosmology}.  
Thus for the remainder of this article, we focus on the specific implementation of the clockwork axion that is defined by \eref{eq:Lagrangian}.  

\section{Cosmological effects of the clockwork axion}
\label{sec:Cosmology}

Clockworking allows the PQ-breaking scale to be much smaller than the axion decay constant \pref{eq:clockworking}, $f_\PQ \ll f_a$.  
Since most studies of axion cosmology assume $f_\PQ \sim f_a$, it is interesting to investigate whether new phenomena can arise when this relation is badly broken.  
(We will see shortly that the topological defect network is qualitatively different from the standard axion cosmology.)  
However, more to the point, the PQ-breaking scale is a threshold for new physics:  the masses of the scalar fields and PQ fermions are all set by $f_\PQ$ times a dimensionless coupling.  
Whether or not these states are accessible to laboratory experiments depends sensitively the value of $f_\PQ$.  
In this section we explore how cosmological evolution constrains the PQ-breaking scale.  

\begin{figure}[t]
\begin{center}
\includegraphics[width=0.95\textwidth]{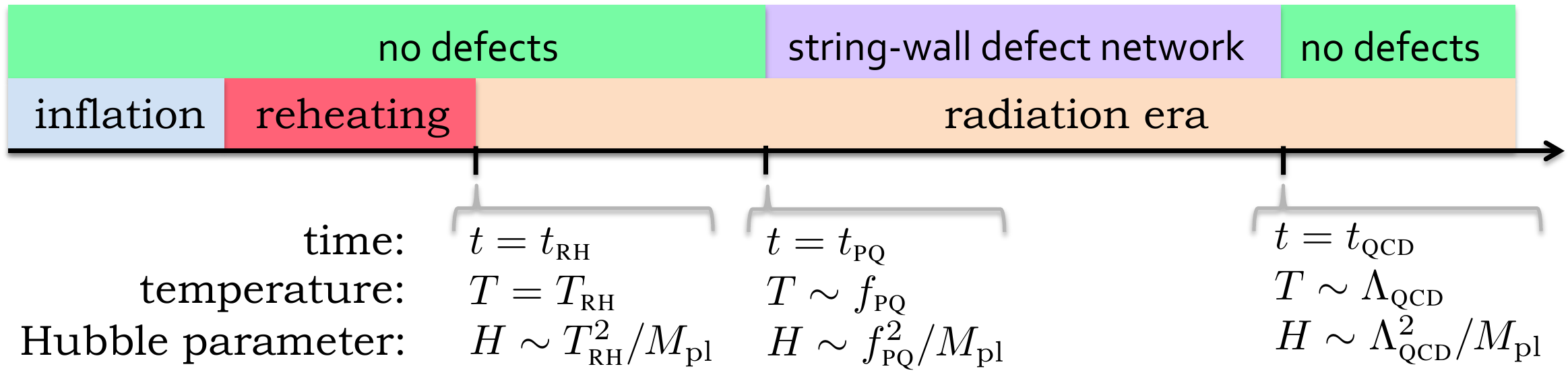} 
\caption{\label{fig:timeline_after}
A timeline illustrating the scenario in which symmetry breaking occurs during the radiation era, after reheating is completed.  The defect network forms at the PQ-breaking phase transition, and it eventually collapses at the QCD epoch when instanton effects lift the axion mass.  Unlike the usual QCD axion cosmology, in the clockwork model the gear-field domain walls form already at $t_\PQ$.  
}
\end{center}
\end{figure}

The scenario we are going to discuss is illustrated in \fref{fig:timeline_after}.  
If $f_\PQ$ is smaller than both the reheating temperature and the Hubble scale during inflation, 
\be
f_\PQ < \mathrm{min}(T_\RH, \frac{H_I}{2\pi}) 
\com 
\ee
and if the fields of \eref{eq:Lagrangian} are in thermal equilibrium with the SM plasma, then a cosmological phase transition occurs when the plasma temperature decreases below $T \sim f_\PQ$.  
During the phase transition the symmetries of the clockwork model become broken, and a network of topological defects forms \cite{Kibble:1976sj}.  
In this section we discuss the rich structure and dynamics of the defect network, and we calculate the axion relic abundance that is emitted from the defect network.  

\subsection{Formation of the defect network}
\label{sec:formation}

Here we are interested in the formation of the defect network at $t = t_\PQ$, and therefore we can neglect the potential induced by QCD instanton effects, which is negligible at temperatures $T \sim f_\PQ \gg \Lambda_\QCD$.  
It is illustrative to first consider the model with $\epsilon = 0$ in \eref{eq:Lagrangian}: in this case the theory admits $N+1$ ``flavors'' of global cosmic strings \cite{VilenkinShellard:1994}.  
We will refer to these strings collectively as $\pi$-strings and individually as $\pi_n$-strings, since we have used $\pi_n/f_\PQ$ to denote the phase of $\phi_n$ in \eref{eq:phi_n}.  
For instance, along a trajectory in spacetime that encircles a $\pi_n$-string with winding number $w \in \Zbb$ the phase of $\phi_n$ evolves smoothly from $0$ to $2\pi w$.  

For $\epsilon\neq 0$ the additional interactions in \eref{eq:Lagrangian} break the $\U{1}^{N+1}$ global symmetry to its $\U{1}_\PQ$ subgroup, which is then spontaneously broken through the cosmological phase transition.  
Naively this suggests that we should only consider one string solution, but in the regime where the $\epsilon$ term can be treated as a perturbation, it is still necessary to consider the $N+1$ $\pi$-strings, since their formation during the cosmological phase transition is unaffected by a perturbatively small explicit symmetry breaking term.  
What differs now is that the $\pi$-strings become bound to one another by domain walls \cite{VilenkinShellard:1994}.  
Along a trajectory through spacetime that crosses a domain wall, one of the gear fields $G_i$ has a nontrivial step-like profile, and therefore we denote this domain wall as a $G_i$-wall.  

In addition to the $N+1$ flavors of $\pi$-strings, the model also admits a global cosmic string, which we call the $a$-string, associated with the spontaneous breaking of $\U{1}_\PQ$.  
Since $\U{1}_\PQ$ is not broken explicitly the $a$-string is topologically stable,\footnote{The vacuum manifold is indeed connected with homotopy group $\pi^1(vac)\neq 1$.} unlike the $\pi$-strings, and also the $a$-string is not connected to any domain walls.  
As one traces a trajectory around the $a$-string in real space, the axion field \pref{eq:clockworking} evolves from $a=0$ to $a=2\pi f_a$ in its configuration space.  
Simultaneously the field $\pi_n$ for $n \in \{ 0, 1, \cdots N \}$ passes through $q^{N-n}$ complete cycles of $\pi_n \in [0, 2\pi f_\PQ)$, which can be seen from \erefs{eq:pi_n}{eq:On0_Oni}.  
In other words, the $a$-string carries the same topological charge as a particular composition of $\pi$-strings that combines an $n$-string with winding number $q^{N-n}$ for each of the $n \in \{ 0, 1, \cdots N \}$.  
As noted in \rref{Higaki:2016jjh}, this observation lets us view the $a$-string as a ``bundle'' of $\pi$-strings that are localized in a region of space that is comparable to the string thickness and much smaller than the inter-string separation.  
See \fref{fig:config_space} to visualize concretely how the string solutions interpolate through the vacuum manifold.  

At the QCD epoch the axion mass is lifted by instanton effects, and the residual $\U{1}_\PQ$ symmetry is broken.  
We assume that the breaking of $\U{1}_\PQ$ leaves no unbroken subgroup ($N_\DW = 1$), and the potential has a single global minimum.  
Consequently the domain wall no longer interpolates between degenerate vacua, but rather the vacuum with lower energy pushes the wall away.  
In this way, the entire string-wall network collapses at the QCD epoch, and any energy it carried is converted into particle emission.  

\subsection{Network of cosmic strings}
\label{sec:strings}

In this section we neglect the presence of the domain walls, and we treat the defect network as $N+1$ independent and non-interacting cosmic string networks, which are distinguished by the strength of their coupling to the axion.  
We are especially interested in the emission of axions from the string network and their contribution to the axion relic abundance \cite{Davis:1986xc,Vilenkin:1986ku,Harari:1987ht}.  

The reader may question whether it is reasonable to neglect the presence of the domain walls, which can pull on the cosmic strings and thereby affect their motions.  
This approach is certainly justified if the gear mass ($\epsilon$ term in \eref{eq:Lagrangian}) arises dynamically at a scale that is much lower than the scale of PQ breaking; for a model that concretely implements this idea, see \rref{Agrawal:2017cmd}.  
However, if the gear mass is already present at the PQ-breaking phase transition, the situation becomes less clear.  
We return to discuss the domain walls in \sref{sec:walls}.  
Nevertheless, in the end we will see that even with this favorable assumption, the axion relic abundance arising from cosmic strings is subdominant to other contributions.  

\subsubsection{Cosmic string solution}
\label{sec:string_solution}

A global string \cite{Vilenkin:1982ks,VilenkinShellard:1994} arises in the theory of a complex scalar field $\phi(x)$ with a symmetry breaking potential $V = \lambda (|\phi|^2 - f^2/2)$.  
The string solution with winding number $w \in \{ \pm 1, \pm 2, \pm 3, \cdots \}$ can be parametrized in cylindrical coordinates as $\phi^{(w)} = (f/\sqrt{2}) \, F(mr) \, \mathrm{exp} (i w \theta)$ where $\langle \phi \rangle = f / \sqrt{2}$ is the vacuum expectation value of $\phi$, $m = \sqrt{2\lambda} f$ is the mass of $\phi$, $r$ is the distance from the center of the string, $\theta$ is the azimuthal angle, $F(x)$ is the radial profile function, and $\delta_s \sim 1 / m$ is the thickness of the string core.  
The profile function satisfies $F(x) = A x^{|w|}$ for $x \ll 1$ and $F(x) \to 1$ for $x \gg 1$ independent of $w$.  

The string tension, $\mu$, measures the string's rest energy per unit length, and it is given by 
\begin{align}\label{eq:mu_def}
	\mu = \pi f^2 \int_0^{mR} \! \ud x \, x \, \Bigl[ F^{\prime}(x)^2 + \frac{w^2}{x^2} F(x)^2 + \frac{1}{4} \bigl( F(x)^2-1 \bigr)^2 \Bigr] 
\per
\end{align}
The second term in the integrand leads to the well-known logarithmic divergence in the tension of global strings.  
It is customary to cut off the radial integral at the typical inter-string separation length scale, $r \sim R$.  
Additionally, it is illustrative to split the integral into two pieces at $x = 1$ corresponding approximately to the edge of the string core at $r = 1/m \sim \delta_s$.  
Specifying to the clockwork axion model, we replace $f \to f_\PQ$ and $m \to m_\rho$.  
Thus the tension of a $\pi$-string with winding number $w$ is estimated as 
\begin{align}\label{eq:mu}
	\mu \approx \mu_{\rm core} + w^2 \, \pi f_\PQ^2 \, \log \bigl( m_\rho R \bigr) 
\end{align}
where $\mu_{\rm core} \approx \pi f_\PQ^2$ arises from $x \in [0, 1)$, and the second, larger term arises from $x \in [1, \infty)$.  
The logarithmic factor can be quite large since the typical inter-string separation is given by the Hubble length scale; for $m_\rho = 1 \TeV$ and $R = 1 / H_\QCD$ we have $\log(m_\rho R) \sim 50$.  

As we discussed in \sref{sec:formation}, the $a$-string can be interpreted as a composition of the various $\pi_n$-strings each with winding number $q^{N-n}$; see also \rref{Higaki:2016jjh}.  
Therefore, \eref{eq:mu} lets us estimate the tension of the $a$-string (with winding number $\pm 1$) to be 
\begin{align}\label{eq:mu_a}
	\mu_a 
	\approx \sum_{n=0}^N \mu \bigr|_{w=q^{N-n}} 
	\approx \pi \, q^{2N} f_\PQ^2 \log \big( m_\rho R \big) 
	\approx \pi f_a^2 \log \big( m_\rho R \big) 
	\per 
\end{align}
where we have used \eref{eq:clockworking} in the last equality.  
Here the approximation requires that the integral of \eref{eq:mu_def} is done in a common region of space for all the string of the ``bundle.''  
Notice that this estimate agrees with the standard expectation for an axion string, $\mu_a \sim f_a^2$ \cite{Vilenkin:1982ks,VilenkinShellard:1994}.  
We can also understand this result by noting that the axion field has to pass through a large field excursion, $a \in [0, 2\pi f_a)$, as we circumnavigate the string, and this large field gradient translates into a large energy density and a large tension.  

\subsubsection{Formation of the $a$-string is prevented}
\label{sec:a_string}

Although the formation of defects depends crucially on initial conditions and dynamics, we argue that -- in the parameter space of interest -- the PQ-breaking phase transition produces a network of $\pi$-strings connected by $G$-walls, and that the $a$-string does not form, neither at the PQ phase transition nor later during the evolution of the network.  

First we argue that the initial conditions at the time of the PQ-breaking phase transition inhibit the formation of the $a$-string.  
During the phase transition the scalar fields $\phi_n$ develop a tachyonic mass of order $m_\rho$, and they acquire vacuum expectation values according to \eref{eq:Lagrangian}.  
In the regime $m_G \ll m_\rho$ where $\U{1}^{N+1}$ becomes a good symmetry, the fields $\phi_n$ evolve almost independently, each one ``seeing'' an identical tachyonic mass of order $m_\rho$, and the defect network that forms consists of $\pi$-strings connected by $G$-walls.  
In the other regime $m_G \sim m_\rho$ where $\U{1}^{N+1}$ is badly broken, the symmetry-breaking potential induces an ``anisotropic'' contribution to the tachyonic mass of $O(m_G)$, which biases the path of the $\phi_n$ toward the vacuum manifold, i.e. the flat direction associated with the axion field, and the resulting defect network consists of $a$-strings without any walls.  
We are interested in the parameter regime where $\epsilon \ll \lambda$ so that $m_G \ll m_\rho$, and we expect that the PQ-breaking phase transition produces a string-wall network rather than a network of $a$-strings.  

Next we argue that the $a$-strings are not formed during the subsequent evolution of the string-wall network.  
Recall that the $a$-string has the same topological charge as a composition of $q^{N-n}$ strings of type $\pi_n$, and for the parameters of interest, $q = 3$ and $N \sim 15$ so that $q^N \sim 10^7$.  
In order to construct an $a$-string from the string-wall network, it is necessary to combine an enormous number of $\pi$-strings.  
Typically there are only $O(1-10)$ long strings in a Hubble volume at any time.  
Therefore, for $N \gg 1$ it is very improbable that the $a$-string will form during the evolution of the string-wall network.  
In fact a numerical simulation of the evolution of the string-wall network appears in \rref{Higaki:2016jjh}.  
For the model with two scalar fields ($N+1=2$) they observe that the $a$-string is able to form from the collapse of the string-wall network, but for the model with three scalar fields ($N+1=3$) the $a$-string does not form, and it is reasonable to expect that formation of the $a$-string is also prevented for models with more than three scalar fields.  

Therefore, in the remainder of this article, we assume that the PQ-breaking phase transition creates a network of $\pi$-strings connected by $G$-walls.  
If a network of $a$-strings were formed instead, then the cosmology would be unchanged from the usual QCD model, and specifically the PQ-scale gear particles would play no role.  
(We assume the gears decay quickly to gluons through the interaction in \eref{eq:piN_GG}.)  

\subsubsection{Evolution of the string network}\label{sec:strings_evolution}

If the $a$-string does not form, as we have argued above, then we can focus on the $\pi$-strings.  
In this section we discuss briefly the dynamical evolution of the string network, and in the next section we estimate the efficiency of axion emission.  

A network of $\pi$-strings forms at the PQ-breaking phase transition.  
Initially particles in the plasma scatter frequently on the strings leading to an effective friction that damps their motion.  
Since the strings interact most strongly with radial modes and gears, and since these particles go out of equilibrium soon after the phase transition, the friction force quickly becomes negligible, and the string network begins to evolve freely under the pull of its tension.  
At this time the network contains at least a few long strings that cross the Hubble volume.  
When long strings intersect one another (or self-intersect), they create large Hubble-scale loops, and subsequently smaller loops are created from the intersections of larger ones.  
After a few Hubble times of free evolution, these dynamics bring the string network into the {\it scaling regime} where new loops are continuously formed at a roughly fixed fraction of the Hubble scale.  
A loop of size $L$ oscillates with a period of approximately $T = L/2$ under the pull of its tension.  
These oscillations cause string segments to be accelerated, which leads to particle emission and gravitational wave radiation.  
The associated energy loss causes the loop to lose energy and shrink.  
Eventually the loop size becomes comparable to its thickness, $\delta_s \sim m_\rho^{-1}$, and it finally decays.  

Then the energy density of the $\pi_n$-string network at time $t$ is given by \cite{VilenkinShellard:1994} 
\begin{align}\label{eq:rho_str}
	\rho_{\rm str}^{(n)}(t) 
	\sim N_{\rm long}^{(n)} \mu H^2 + \frac{\mu H^2}{\sqrt{ P^{(n)} / \mu}}
\end{align}
where the first term arises from the Hubble-scale long strings, and the second term arises from the string loops.  
Here $N_{\rm long}^{(n)}$ is the number of long strings, $\mu$ is the string tension \pref{eq:mu}, $H = H(t)$ is the Hubble parameter at time $t$, and $P^{(n)}$ is the average power emitted by the loop during each oscillation period.  
The derivation of \eref{eq:rho_str} assumes that $P^{(n)}$ is independent of $t$ and the loop length $L$.  
In general $N_{\rm long}^{(n)}$ and $P^{(n)}$ can be different for the various flavors of $\pi_n$-strings, whereas $\mu$ is universal.  
We expect that $N_{\rm long}^{(n)}$ is either zero or $O(1)$, because the ``chopping'' of long strings prevents $N_{\rm long}^{(n)} \gg 1$.  
There is a model-independent contribution to the power from gravitational wave emission, which is parametrically $P^{(n)} \sim G_N \mu^2$ with $G_N$ Newton's constant, but the emission of Goldstone bosons typically dominates as we discuss in the next section.  
One can understand the enhancement factor, $(P^{(n)}/\mu)^{-1/2} \geq 1$, because loops that do not emit efficiently will live longer and contribute to $\rho_{\rm str}$.  

\subsubsection{Axion emission from strings}\label{sec:strings_emission}

Let us now consider the emission of axions as the string network evolves from the PQ-breaking phase transition to the QCD epoch.  
It is useful to first recall the axion emission calculation for standard QCD axion strings.  

For the usual QCD axion, one estimates the energy density in axions at time $t$ as $\rho_a(t) \sim \varepsilon \times \mu_a H^2 \times \log t/t_\PQ$, for example by requiring self-consistency with the scaling solution \cite{Harari:1987ht,VilenkinShellard:1994,Hagmann:1998me,Hiramatsu:2010yu}.  
The first factor, $\varepsilon \sim P / \mu \sim O(1)$, accounts for the high efficiency with which axion strings emit the pseudo-Goldstone axions.  
The second factor, $\mu_a H^2$, is the total energy density in the network of strings with tension $\mu_a$ that are in the scaling regime \pref{eq:rho_str}.  
The logarithmic factor accounts for integrating the axion emission from time $t_\PQ$ to time $t$.  
In particular, note that $\rho_a \propto \mu_a \sim f_a^2$.  

By translating these results to the QCD clockwork axion model, we can immediately infer the axion abundance.  
For models with very small explicit breaking, $\epsilon \ll 1$ such that $L \ll m_G^{-1}$, each $\pi_n$-string will very efficiently emit their corresponding pseudo-Goldstone boson, $\pi_n$, and we expect an energy density $\rho_{\pi_n} \propto \mu H^2$ where $\mu \sim f_\PQ^2$ is the tension of the $\pi_n$-string.  
Since the axion $a$ is a linear combination of the $\pi_n$'s, we also expect that the axion energy density is 
\begin{align}\label{eq:rhoaQCD_string}
	\rho_a(t_\QCD) \sim \mu H_\QCD^2 \, \log t_\QCD / t_\PQ
	\per
\end{align}
Since $\mu \sim f_\PQ^2 \sim q^{-2N} f_a^2$ from the clockworking, the predicted axion abundance is exponentially suppressed with respect to the usual QCD axion case.  
(We will see in \sref{sec:relics} that the corresponding axion dark matter is also negligible compared to the misalignment contribution.)  

The preceding analysis assumes that the gear mass is negligible compared to the loop size, which implies that string loops efficiently radiate all the $\pi_n$.  
The more realistic scenario corresponds to the opposite regime, $m_G^{-1} \ll L$, and here the calculation of $\rho_a$ is more subtle, because the emission of massive gears, $G_i$, is exponentially suppressed, and the emission of the massless axion, $a$, depends on the strength of its coupling to each different flavor of $\pi_n$-string.  
Since the loops do not radiate so efficiently, they also live longer, and they carry more energy at late times [$P^{(n)}/\mu \ll 1$ in \eref{eq:rho_str}].  
These effects parametrically enhance the axion relic abundance, but nevertheless the enhancement does not compensate the exponential suppression, $f_\PQ^2 / f_a^2 \sim q^{-2N} \ll 1$, and the axion emission from the string network is still much smaller than the standard QCD axion result.  

\subsection{Network of gear-field domain walls}
\label{sec:walls}

In this section we discuss the network of domain walls that connects the $\pi$-strings, which are built from the gear fields, and we calculate the relic abundance of axions emitted by the walls.  

\subsubsection{Domain wall solution}
\label{sec:wall_solution}

In general domain wall solutions arise in theories for which the scalar potential has two or more minima separated by barriers \cite{VilenkinShellard:1994}.  
If the potential can be approximated by a quartic polynomial $V = (\lambda/4) (\varphi^2 - v^2)^2$, then the domain wall solution is parametrized as $\varphi = \pm v \, \tanh (m z / 2)$ where $z$ is the spatial coordinate normal to the wall (assumed planar), $\pm$ is for kink and anti-kink solutions, and $\delta_w \sim 1 / m$ is the wall thickness, which is related to the mass $m = \sqrt{2 \lambda} \, v$.  
The wall's surface tension $\sigma$ measures its energy per unit area and for this model the tension is $\sigma = \sqrt{8/9} \, m v^2$.  
Alternatively, if the potential is written as $V = m^2 f^2 \, \cos( \varphi / f)$ then the domain wall solutions take the form $\varphi = \pm \bigl( 4 f \, \arctan \, e^{m z} - \pi f \bigr)$ where $\delta_w \sim 1/m$ and $\sigma = 8 m f^2$.  

For the clockwork axion model, we set $f \to f_\PQ$ and $m \to m_G$ to estimate the surface tension of the $G$-walls as 
\begin{align}\label{eq:sigma}
	\sigma \sim 8 \, m_G f_\PQ^2 
	\com 
\end{align}
and the wall thickness is $\delta_w \sim 1 / m_G$.  
(Actually, since the gear spectrum is split \pref{eq:mGi} each wall will have a different tension and width, but this feature does not play a significant role in the dynamics, and we simply use \eref{eq:sigma} to estimate the tension for all of the walls.)  

\begin{figure}[t]
\begin{center}
\includegraphics[width=0.46\textwidth]{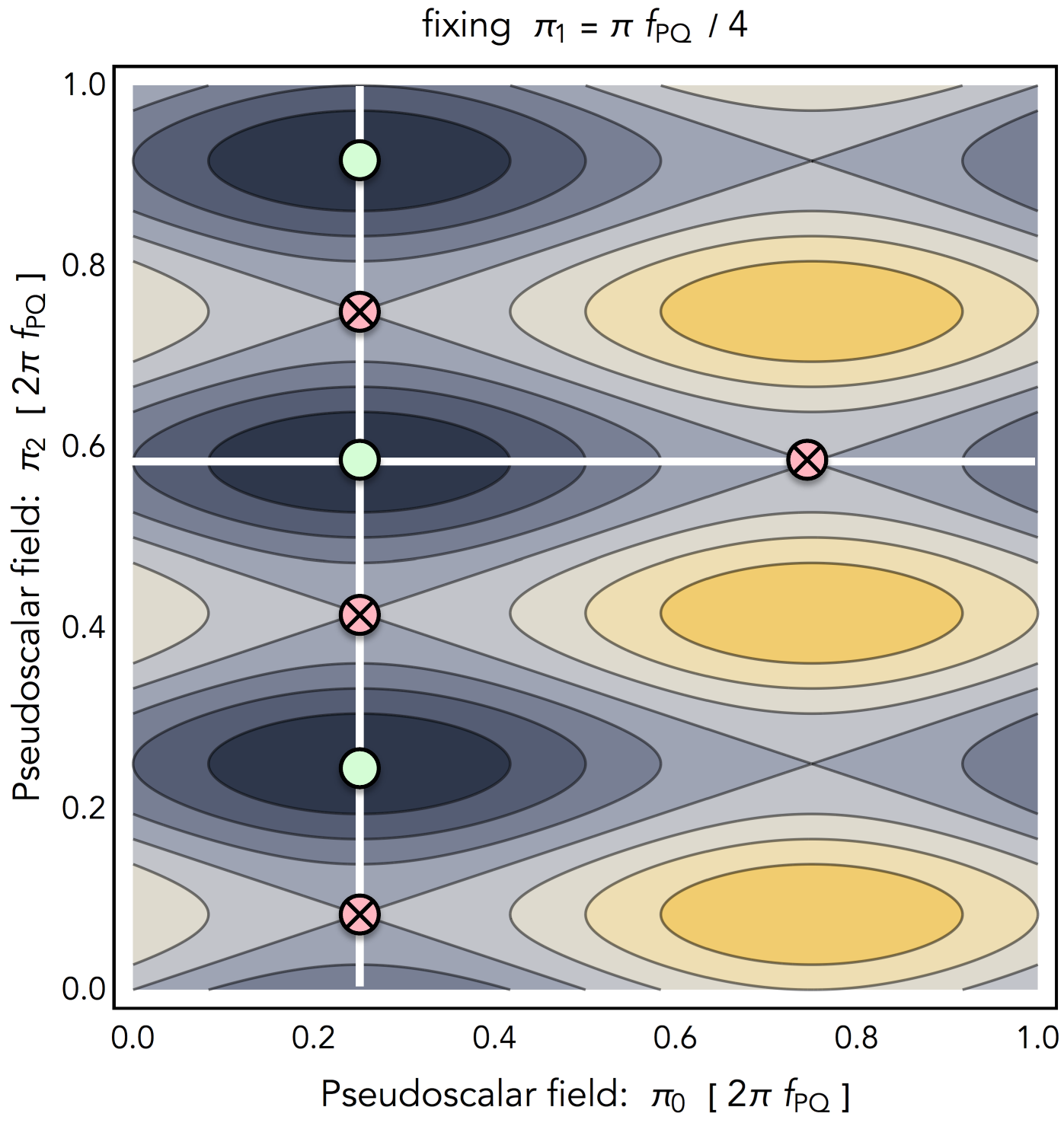} \hfill 
\includegraphics[width=0.46\textwidth]{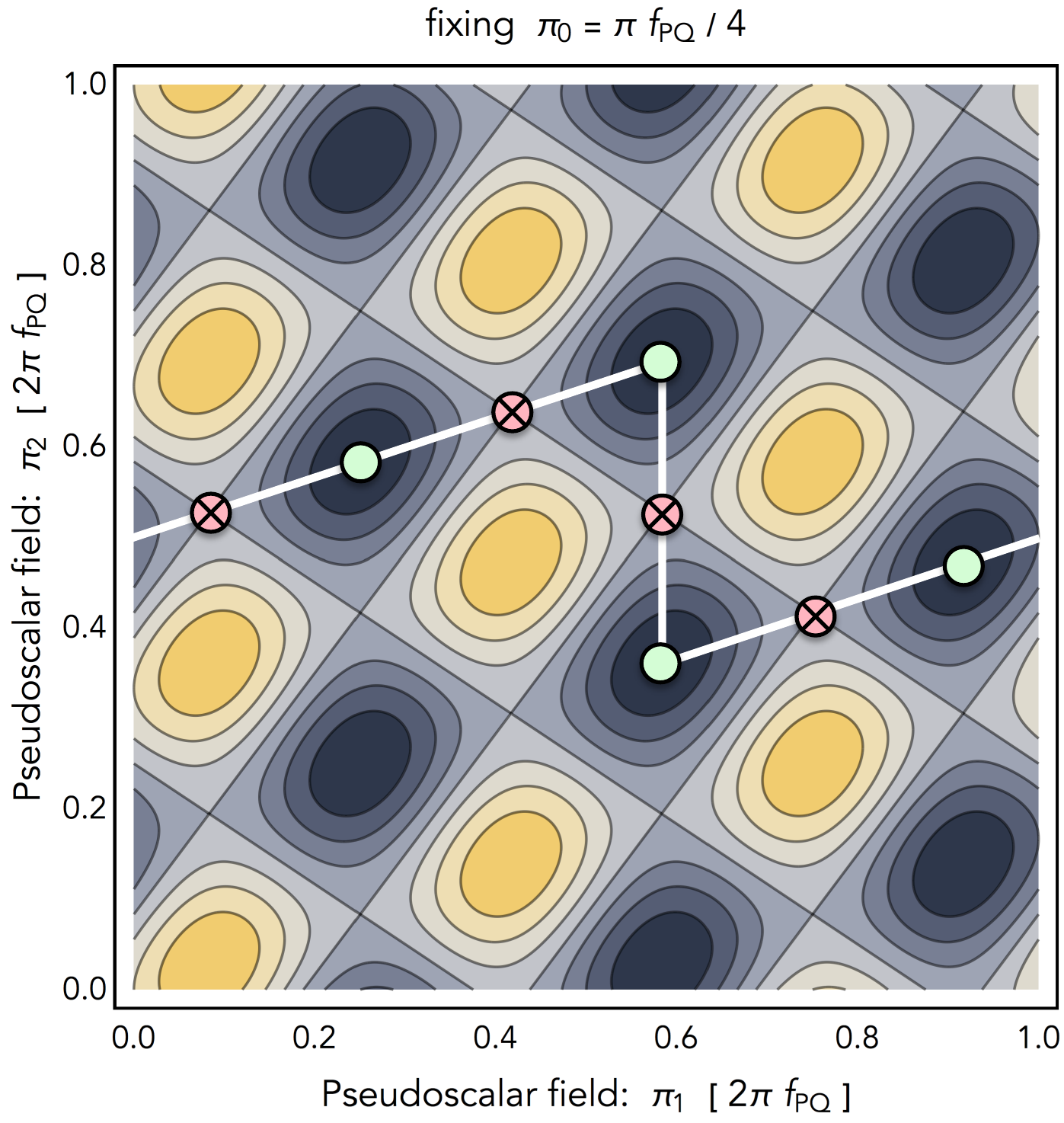} 
\caption{\label{fig:wall_potential}
For the model with three pseudoscalar fields ($\pi_0$, $\pi_1$, and $\pi_2$) we show the scalar potential over slices of the field space, and we overlay the trajectories corresponding to the $\pi_0$-, $\pi_1$-, and $\pi_2$-strings.  The potential is $V = m_G^2 f_\PQ^2 \bigl( \cos [ \theta_0 - 3 \theta_1 ] + \cos [ \theta_1 - 3 \theta_2 ] \bigr)$ where $\theta_i \equiv \pi_i / f_\PQ$.  As we encircle a $\pi_0$-string in spacetime, the fields pass along the horizontal white line (left panel) from the local minimum (green dot) to the saddle point (red cross), and back to the local minimum; therefore, the $\pi_0$-string is connected to $1$ domain wall.  Similarly the $\pi_2$-string, which corresponds to the vertical white line (left panel), connects to $3$ domain walls, and the $\pi_1$-string, which is shown on the right panel, connects to $4$ domain walls.  This discussion readily generalizes to $N+1 > 3$.  (The ``local minima'' are all connected by a flat direction, which is the axion field.)
}
\end{center}
\end{figure}

A given $G_i$-wall in the network can either form a closed bubble, or it can terminate at one of the $\pi_n$-strings.  
Moreover, the $\pi_n$-strings may connect to multiple $G$-walls.  
Let $N_\DW^{(n)}$ denote the number of $G$-walls connected to a $\pi_n$-string.  
From the structure of the interactions in \eref{eq:Lagrangian}, we see that 
\begin{align}\label{eq:num_walls}
	N_\DW^{(n)} = \begin{cases} 1 & , \ n = 0 \\ 4 & , \ n \in \{ 1, 2, \cdots N-1 \} \\ 3 & , \ n = N \end{cases} 
	\per
\end{align}
This situation is illustrated in \fref{fig:wall_potential}, and the caption provides further explanation; see also \rref{Higaki:2016jjh}.  

\subsubsection{Evolution of the wall network}\label{sec:walls_evolution}

Given the rich structure of the clockwork axion string-wall defect network, it is challenging to study its evolution, either analytically or numerically.  
However, we gain some understanding from simple physical arguments.  
Since strings are pulled by multiple walls with different tensions \pref{eq:sigma}, we expect that the string-wall network will begin to collapse, even before the QCD epoch.  
For example, consider a particular string that is connected to several walls.  
It is energetically preferable for the string to move in a direction that shortens the wall with the largest tension.  
In other words, the high-tension walls start to pull together the pair of strings that they are connecting.  
This attractive force accelerates the string segment, which causes it to radiate as we discussed in \sref{sec:strings}.  
The energy loss into radiation causes the string-wall system to shrink.  
Eventually the strings meet, and they either annihilate (if they had opposite winding number) or they merge into a new hybrid string with a composition of the winding numbers of the constituent $\pi$-strings.  
In this way, we expect that the string-wall network will partially collapse.  

This dynamical behavior has recently been observed in numerical simulations of aligned (clockwork) axion models with $2$ and $3$ complex scalar fields.  
For the model with two scalar fields ($N+1=2$), the simulations of \rref{Higaki:2016jjh} show that the string-wall network completely collapses to form a network of hybrid strings, which are precisely the $a$-string that we discussed in \sref{sec:formation}.  
For model with three scalar fields ($N+1 = 3$), the simulations suggest that a partial collapse will occur in which the $\pi_0$-strings, which connect to only a single wall, are merged with the $\pi_1$-strings to form hybrid strings, but these hybrid strings do not merge with the $\pi_2$-strings.  

It is difficult to assess how far does the partial collapse proceed before the system reaches a stationary scaling behavior.  
Surely the network cannot completely collapse for models with many fields, $N \gg 1$, because this would require a network of $a$-strings to form, but we have already argued in \sref{sec:a_string} that this does not occur.  
Therefore, we expect that the partial collapse will suppress the energy carried by $\pi_n$-strings and their connected walls with low $n \lesssim {\rm few}$, and we assume that the network evolution is unaffected by the partial collapse for $n \gg 1$.  
Since our primary interested is to calculate the axion relic abundance, we do not expect more than a factor of $O(N) \sim 10$ uncertainty.  

After the partial collapse has occurred, we expect that the subsequent evolution of the string-wall network is similar to the string-wall network evolution in a QCD axion model with $N_\DW > 1$ \cite{Hiramatsu:2012sc}.  
Specifically, we assume that the network reaches the scaling regime where new walls are created from the ``chopping up'' of larger walls.  
In the scaling regime, we estimate the energy density of the wall network at time $t$ as \cite{VilenkinShellard:1994}
\begin{align}\label{eq:rho_walls}
	\rho_{\rm walls}(t) \sim \sum_{i} \sigma_i H
\end{align}
where $\sigma_i \sim m_{G_i} f_\PQ^2$ is the tension of the $G_i$-wall \pref{eq:sigma}, and $H(t)$ is the Hubble parameter.  
Since the energy of a wall is proportional to its area, the energy density of the network is dominated by the largest Hubble-scale walls, which makes this estimate more robust.  

\subsubsection{Axion emission from walls}
\label{sec:walls_emission}

In this section we estimate the energy density of axions that are emitted from the network of $G$-walls.  
A formalism for calculating particle emission from domain walls has been developed in \rref{Vachaspati:1984yi}, and we loosely follow that approach here.\footnote{The emission of axions from axion-walls has been considered previously in various references, including Refs.~\cite{Hagmann:1990mj,Nagasawa:1994qu,Chang:1998tb,Hagmann:2000ja}, where it is generally assumed that an $O(1)$ fraction of the energy density carried by the domain walls goes into non-relativistic axions.  We cannot apply this assumption to calculate axion emission from the gear-walls, because we expect that most of the energy goes into gears rather than axions.  }    
(The reader may also recall the calculation of electromagnetic radiation from an oscillating sheet of charge, which follows a similar approach.)  
We first identify the interaction between the wall-forming field and the radiation, which correspond to the gear $G_i$ and the axion $a$.  
Next we assume a profile for the $G_i$ and solve the field equation for $a$ to determine the radiated energy.  

Let us inspect the Lagrangian \pref{eq:Lagrangian} to identify the axion-gear coupling.  
Since the axion has an exact shift symmetry before the QCD epoch, interactions with the other pseudoscalar particles can arise in general only through higher derivative operators.
By integrating out the $n^{\rm th}$ radial mode, $\rho_n$, one generates the operator $(\partial \pi_n)^4$.  
We write this operator in the mass basis using \eref{eq:pi_n}, and we focus on  the single-axion term, which is 
\begin{align}\label{eq:J_walls}
	\mathscr{L}_{\rm int}
	& \supset \frac{1}{f_a} \partial_\mu a \, J_{\rm walls}^\mu
	\qquad \text{with} \qquad 
	J_{\rm walls}^\mu \equiv \frac{f_a}{2 f_\PQ^2 m_\rho^2} \, \sum_{n=0}^N\sum_{i,j,k=1}^N b_{n,ijk} \, \partial_\mu G_i \partial^\nu G_j \partial_\nu G_k 
	\per
\end{align}
The coefficient\footnote{The term with two axion and two gear fields has the coefficient $c_{n,ij} = 6 O_{n0} O_{n0} O_{ni} O_{nj} \sim q^{-2n} N^{-1}$.  For $n \gg 1$ this term is suppressed by an additional factor of $q^{-n}$ compared to the single axion term in \eref{eq:J_walls}.  Even for $n \sim 1$ we expect that the two axion term will give a subdominant contribution to the axion emission calculation \cite{Long:2014mxa}.} is defined by $b_{n,ijk} \equiv 4 O_{n0} O_{ni} O_{nj} O_{nk}$, and using \eref{eq:On0_Oni} we estimate $b_{n,ijk} \sim q^{-n} N^{-3/2}$.  
In the approximation where the single axion term is the dominant one, the axion field equation becomes 
\begin{align}\label{eq:Box_a}
	\Box a = \frac{1}{f_a} \partial_\mu J_{\rm walls}^\mu
	\per 
\end{align}
In the vicinity of a $G$-wall the right side of \eref{eq:Box_a} becomes a source for the axion field, and the associated axion radiation is given by the solution evaluated far away from the wall.  

In order to solve for the axion emission we must first know how the gear fields are evolving.  
In principle a numerical simulation of the string-wall network evolution can measure $G_i(x)$ and solve \eref{eq:Box_a} directly to determine the amount of axion emission.  
Since this technique is not available to us, we take a semi-analytical approach instead, and we comment on the main sources of uncertainties at the end.  

Without detailed information on the network evolution, we need to make an ansatz for the trajectory of the domain wall; this is an essential ingredient, because a static domain wall does not radiate.  
In the cosmological setting, the wall will be accelerated by the pull of its own tension and the tension of the strings to which it connects.  
Motivated by the stationary domain wall solution that was discussed in \sref{sec:wall_solution}, we parametrize an accelerated wall by writing\footnote{As we discussed in \sref{sec:wall_solution}, a $\tanh$ profile is the domain wall solution for a double-well quartic polynomial potential.  The gear potential is more accurately given by a sinusoidal potential, which has instead $\arctan \, {\rm exp}$ as its domain wall solution.  We do not expect that the detailed shape of the domain wall profile will significantly impact our results,  and therefore we choose to work with the $\tanh$ profile for convenience.  }
\begin{align}
	G_i \sim \bar{G}(t,x,z) \equiv \pi f_\PQ \, \tanh \Bigl[ m_G \bigl( z - z_0(t,x) \bigr) / 2 \Bigr]
	\com 
\end{align}
which represents a wall located at $z = z_0(t,x)$ with width $\delta_w \sim 1 / m_G$.  
The wall is not planar, but rather it has oscillatory (standing wave) perturbations that can be written as 
\begin{align}\label{eq:z0_param}
	z_0(t,x) = A_\omega \, \sin(\omega x) \, \sin(\omega t) 
\end{align}
where $A_\omega(t)$ is the displacement amplitude at time $t$.  
In the vicinity of this domain wall, the axion field evolves subject to 
\begin{align}\label{eq:a_field_eqn}
	\Box a = \Scal_a
	\quad \text{with} \quad 
	\Scal_a = \lambda_G \, \partial_\mu \bigl( \partial^\mu \bar{G} \partial^\nu \bar{G} \partial_\nu \bar{G} \bigr) 
	\quad \text{and} \quad 
	\lambda_G \equiv \frac{1}{2 f_\PQ^2 m_\rho^2} \sum_{n=0}^{N} b_{n,111} 
	\per 
\end{align}
Parametrically, the effective coupling strength is $\lambda_G \sim (N^{3/2} f_\PQ^2 m_\rho^2)^{-1}$ since the $n=0$ term dominates the sum over $n$.  

Formally the solution of \eref{eq:a_field_eqn} is written as 
\begin{align}\label{eq:a_soln}
	a(x) = \int \! \ud^4 x^\prime \, \Delta_R(x,x^\prime) \, \Scal_a(x^\prime)
\end{align}
where $\Delta_R(x,x^\prime)$ is the retarded Green's function.  
It is now a straightforward (but lengthy!) exercise to evaluate this integral.  
To determine the amount of axion radiation, we focus on the regime $z \to \infty$ where we find\footnote{We could have guessed the parametric behavior seen in this solution by studying \eref{eq:Box_a} directly by writing $\partial_t G \sim (m_G A_\omega \omega) G$ and $\partial_z G \sim m_G G$ such that $\Scal_a \sim \lambda_G \partial(\partial G)^3 \sim \lambda_G \partial_t^2 \partial_z^2 G^3 \sim \lambda_G (m_G A_\omega \omega)^2 m_G^2 (\pi f_\PQ)^3$.  Then  $\Box a \sim (\partial_t^2 - \partial_z^2) a \sim m_G^2 a$.  }
\begin{align}\label{eq:a_far_field}
	a \sim \lambda_G \, \pi^3 f_\PQ^3 \, m_G^2 \, A_\omega^2 \, \omega^2 \, \cos (2 \omega t - 2 \omega z)
\end{align}
up to an $O(1)$ numerical coefficient.  
(The field amplitude approaches a constant far from the wall, as in the case of electromagnetic radiation from a sheet of charge, and unlike the case of radiation from a point charge, which gives instead a familiar $1/r$ behavior.)  

As the wall oscillates it radiates axions with a energy flux $T_{03}$, which has the units of power per area.  
The energy flux is estimated as 
\begin{align}\label{eq:energy_flux}
	T_{03} 
	= \langle \partial_t a \, \partial_z a \rangle 
	\sim \omega^2 a^2 
	\sim \lambda_G^2 \, \pi^6 f_\PQ^6 \, m_G^4 \, A_\omega^4 \, \omega^6
\end{align}
where the angled brackets indicate time averaging over one oscillation period.  
We see that axion emission is more efficient for walls with high frequency oscillations, which follows from the derivative interaction between $a$ and the $G_i$.  

The oscillating domain wall emits axions at the expense of reducing its own kinetic energy and decreasing the amplitude of its oscillations.  
The domain wall's kinetic energy per unit area is $(1/2)\sigma A_\omega^2 \, \omega^2$, since $A_\omega \, \omega$ is the effective oscillation velocity and the tension  $\sigma \sim m_G f_\PQ^2$ is the mass per unit area.  
Axion emission leads to 
\begin{align}
	\frac{d}{dt} \Bigl[ \frac{1}{2} \sigma A_\omega^2 \, \omega^2 \Bigr] = - T_{03}
	\com
\end{align}
which gives the evolution of $A_\omega(t)$.  
Using the expression for $T^{03}$ from \eref{eq:energy_flux}, the solution is 
\begin{align}\label{eq:Asol}
	A_\omega(t) 
	= A_\omega(t_1) \Bigl( 1 + 2 \sigma^{-1} v_\omega^2 \lambda_G^2 \, \pi^6 f_\PQ^6 m_G^4 \omega^2 \bigl( t - t_1 \bigr) \Bigr)^{-1/2} 
\end{align}
where $A_\omega(t_1)$ is the initial amplitude of the mode with frequency $\omega$ at an arbitrary time $t_\PQ \leq t_1 < t$.  
We have also defined the dimensionless parameter $v_\omega \equiv A_\omega(t_1) \omega$, which represents the initial speed of the oscillation with frequency $\omega$.  
From \eref{eq:Asol} we see that the mode with frequency $\omega$ has a ``lifetime'' given by 
\begin{align}
	\tau_\omega \sim \Bigl( 2 \sigma^{-1} v_\omega^2 \lambda_G^2 \, \pi^6 f_\PQ^6 m_G^4 \omega^2 \Bigr)^{-1} 
	\per
\end{align}
For $t - t_1 < \tau_\omega$ the mode amplitude is effectively constant, but for $t - t_1 > \tau_\omega$ the mode amplitude decreases like $A_\omega \sim 1 / \sqrt{t}$ as axion emission damps the wall's oscillations.  

The initial oscillation spectrum, parametrized by $A_\omega(t_1)$ or $v_\omega = A_\omega(t_1) \omega$, represents the largest source of uncertainty in our calculation.  
A calculation of this spectrum is not analytically tractable, given the chaotic and dynamic nature of the string-wall network.  
The two essential questions are: how does $A_\omega(t_1)$ depend on $\omega$ and what it the magnitude of $A_\omega(t_1)$?  
To address these questions, we simply assume that as modes on the domain wall enter the horizon, they have a displacement amplitude that is some fraction $\alpha < 1$ of the Hubble scale at that time.  
In other words, the mode with frequency $\omega$ enters the horizon at time $t_\omega$ such that 
\begin{align}
	t_\omega \sim d_H(t_\omega) \sim 1 / \omega
	\qquad \text{and} \qquad 
	A_\omega(t_\omega) \sim \alpha \, d_H(t_\omega) \sim \alpha / \omega
	\per 
\end{align}
In effect this assumption lets us treat the product $v_\omega = A_\omega(t_\omega) \, \omega \sim \alpha \leq 1$ as a constant that is static and independent of $\omega$.  
We do not attempt to calculate $\alpha$ from first principles, but we have no reason to expect that it should not be $O(1)$.  

From \eref{eq:Asol} we identify a characteristic frequency, 
\begin{align}\label{eq:omega_ast}
	\omega_\ast(t) \equiv \frac{1}{\sqrt{2} \sigma^{-1/2} v_\omega \lambda_G \, \pi^3 f_\PQ^3 \, m_G^{2} \, \sqrt{t}} 
	\per 
\end{align}
This characteristic frequency sets a smoothing length scale $l_\ast(t) = 1 / \omega_\ast(t)$ that is important for understanding the structure of the domain wall and its axion emission.  
If we consider the spectrum of oscillations on the domain wall at time $t$, the high-frequency (small-scale) modes with $\omega > \omega_\ast(t)$ are absent, because they were damped away at earlier times due to axion emission, and the low-frequency (large-scale) modes with $\omega < \omega_\ast(t)$ are still present, because they do not yet emit axions efficiently.  
Therefore the modes with $\omega = \omega_\ast(t)$ will give the dominant contribution to the axion emission at time $t$.  
At the QCD epoch we have $\omega_\ast(t_\QCD) \sim {\rm MeV} / v_\omega$ for $N=15$, $m_\rho = 10 \TeV$, and $m_G = 100 \GeV$.  
These modes are much smaller than the horizon scale ($H_\QCD / \omega_\ast(t_\QCD) \sim 10^{-16} v_\omega$), but also much larger than the wall thickness ($m_G / \omega_\ast(t_\QCD) \sim 10^3 v_\omega$).  

Collectively the domain walls form a network that radiates axions with a power density $\Pcal_a(t)$, which has the units of power per volume.  
As we saw in \sref{sec:walls_evolution}, the horizon-scale domain walls carry most of the energy, and we can estimate the power density as $\Pcal_a(t) \sim T_{03} H$.  
The energy density of radiated axions satisfies 
\begin{align}\label{eq:rhoa_eqn}
	\dot{\rho}_a + 4 H \rho_a = \Pcal_a
\end{align}
where the term $4 H \rho_a$ appears because the axions are massless prior to the QCD epoch.  
We integrate the evolution equation for $\rho_a$ to find the energy density of emitted axions at the QCD epoch.  
During the radiation-dominated era we have $H(t) = 1/(2t)$, and the solution is 
\begin{align}\label{eq:rhoaQCD_wall}
	\rho_a(t_\QCD) \sim v_\omega^2 \Biggl[ \frac{\omega_\ast^2}{\omega^2} \log\bigl( 1 + \frac{\omega^2}{\omega_\ast^2}\bigr) - \frac{1}{1 + \omega^2 / \omega_\ast^2} \Biggr] \sigma H_\QCD
\end{align}
where $\omega_\ast$ is evaluated at $t_\QCD$, and $\sigma \sim m_G f_\PQ^2$ is the wall's surface tension.  
\eref{eq:rhoaQCD_wall} is a main result of this work.  

Inspecting \eref{eq:rhoaQCD_wall}, we see that the quantity in square brackets peaks at $\omega / \omega_\ast \simeq 1.5$ where its value is $0.22$.  
If the wall is oscillating with a low frequency, $\omega \ll \omega_\ast(t_\QCD)$, then there has not yet been enough time for axion emission to occur, and $\rho_a$ is suppressed like $(\omega/\omega_\ast)^2$.  
On the other hand, if $\omega \gg \omega_\ast(t_\QCD)$ then the axion emission has occurred long before the QCD epoch, and redshifting has diluted the axion energy density leading to a suppression of $(\omega / \omega_\ast)^{-2}$.  

In the Introduction we asserted that $\rho_a(t_\QCD) \sim \varepsilon^\prime m_G f_\PQ^2 H_\QCD$ \pref{eq:Oah2_naive}, and now by comparing with \eref{eq:rhoaQCD_wall} we identify the efficiency factor as 
\begin{align}\label{eq:epsilon_prime}
	\varepsilon^\prime 
	\sim v_\omega^2 \Biggl[ \frac{\omega_\ast^2}{\omega^2} \log\bigl( 1 + \frac{\omega^2}{\omega_\ast^2}\bigr) - \frac{1}{1 + \omega^2 / \omega_\ast^2} \Biggr]
	\sim \begin{cases}
	0.1 v_\omega^2 & , \quad \omega \sim \omega_\ast(t_\QCD) \\ 
	\frac{\pi^6}{N^3} v_\omega^4 \frac{m_G^4}{m_\rho^4} \frac{\omega^2}{m_G H_\QCD} & , \quad \omega \ll \omega_\ast(t_\QCD)
	\end{cases}
\end{align}
where we have used $\sigma \sim m_G f_\PQ^2$.  
For the usual QCD axion, the dark matter relic abundance is dominated by axion emission from Hubble-scale domain wall oscillations at the QCD epoch.  
To evaluate this contribution in the clockwork axion model, we evaluate the efficiency factor with $\omega \sim H_\QCD$ to find that $\varepsilon^\prime$ is suppressed by $(H_\QCD / m_G) \ll 1$. 
Therefore we can anticipate the result that appears in \sref{sec:relics}; namely, the axion dark matter emission from oscillating gear-field domain walls is negligible.  
On the other hand, in the formula for energy flux \pref{eq:energy_flux} we have seen that high frequency oscillations on the domain wall lead to a larger axion emission.  
For the modes with $\omega \sim \omega_\ast(t_\QCD)$ we find $\rho_a(t_\QCD) \sim 0.1 v_\omega^2 \sigma H_\QCD$, which corresponds to extracting an $O(0.1$) fraction of the wall's kinetic energy as axion emission.  
However these axions are highly boosted since $E_a(t_\QCD) \sim \omega_\ast(t_\QCD) \gg H_\QCD$, and they remain relativistic during the epoch of recombination.  
Therefore they should be treated as a dark radiation, rather than a dark matter.  

\subsection{Collapse of the string-wall network}
\label{sec:collapse}

The axion mass arises at the QCD epoch when instanton effects induce a potential for $\pi_N$; see \eref{eq:piN_GG}.  
Provided that $N_{DW} = 1$, the flat directions are lifted leaving only a single, unique vacuum.  
As a result the string-wall network collapses, and the energy is liberated as mostly gears and axions.  
Here we estimate the amount of energy that goes into the axion.  

The instanton-induced axion potential causes a mixing between the gear fields and the axion.  
The mixing with the $i^{\rm th}$ gear is estimated as 
\be\label{eq:mixing}
\sin \theta_i \sim \frac{q^N}{C} \frac{m_a^2}{m_{G_j}^2} 
\ee
for $i \in \{ 1, 2, \cdots, N \}$.  
Despite the enhancement from $q^N \gg 1$, this factor is typically very small due to $m_a^2 \ll m_{G}^2$.  

The mixing \pref{eq:mixing} controls the efficiency with which the defect network can emit axions.  
The total energy density in the defect network at the QCD epoch is dominated by the Hubble-scale domain walls, which we estimate with \eref{eq:rho_walls} to be $\sim \sigma H_\QCD$ where $\sigma \sim m_G f_\PQ^2$ is the surface tension.  
Therefore the axion energy density that arises from the collapsing string-wall network at the QCD epoch is estimated to be 
\begin{align}\label{eq:rhoaQCD_collapse}
	\rho_a(t_\QCD) 
	\sim \sum_{j=1}^{N} \theta_j^2 \, \sigma H_\QCD 
	\sim N \frac{m_a^4 f_a^2}{m_G^3} H_\QCD 
	\per
\end{align}
Here we have used \eref{eq:clockworking} to write $q^N f_\PQ \sim f_a$.  
Since the usual QCD axion gives $\rho_a \sim m_a f_a^2 H_\QCD$, we see that the axion emission here is suppressed by $(m_a/m_G)^3$.  
The axions emitted from Hubble-scale walls are expected to have an energy $E_a \sim H_\QCD$ so that they become nonrelativistic soon after the QCD epoch \cite{Hiramatsu:2012gg}.  
The remainder of the energy is emitted mostly into gears $G_i$ that decay to Standard Model radiation through their interaction with gluons \pref{eq:piN_GG}.  
The associated entropy injection heats the SM plasma momentarily, but since the energy of the defect network is negligible compared to the energy of the plasma, the heating is not significant.  

\subsection{Axion relic abundance}
\label{sec:relics}

In this section we calculate the relic abundance of dark matter and dark radiation in the clockwork axion model.  
In the subsections, we enumerate the various contributions to $\rho_a(t_\QCD)$.  

The population of nonrelativistic axions contributes to the axion dark matter relic abundance, which is parametrized by $\Omega_a h^2 = \rho_a(t_0) / 3 \Mpl^2 \bar{H}_0^2$ where $\bar{H}_0 \equiv 100 \km / {\rm sec} / {\rm Mpc}$ and $\rho_a(t_0)$ is the energy density of nonrelativistic axions today.  
Between the QCD epoch and today the energy density of nonrelativistic axions redshifts as $\rho_a(t_0) = \rho_a(t_\QCD) \bigl( a(t_0) / a(t_\QCD) \bigr)^{-3}$.  
Assuming adiabatic expansion of the cosmological plasma from the QCD epoch until today, the axion relic abundance is written as 
\begin{align}\label{eq:Oah2_analytic}
	\Omega_a h^2 = \frac{\rho_a(t_\QCD)}{3 \Mpl^2 \bar{H}_0^2} \left( \frac{g_{\ast S}(t_0) T_0^3}{g_{\ast S}(t_\QCD) T_\QCD^3} \right) 
	\simeq \bigl( 4 \times 10^9 \GeV^{-4} \bigr) \rho_a(t_\QCD)
\end{align}
where $g_{\ast S}(t_\QCD) \simeq 20$, $T_\QCD \simeq 0.2 \GeV$, $g_{\ast S}(t_0) \simeq 3.91$, and $T_0 \simeq 0.234 \meV$.  
The measured dark matter relic abundance is $\Omega_\DM h^2 \simeq 0.12$ \cite{Planck:2015xua}.  

The population of relativistic axions contributes to the axion dark radiation relic abundance, which is parametrized by $\Delta N_{\rm eff}(t) = \rho_a(t) / [ 2 (7/8) (\pi^2/30) (4/11)^{4/3} T(t)^4 ]$.  
The presence of a dark radiation during the epoch of recombination ($t = t_{\rm rec}$) is strongly constrained by observations of the cosmic microwave background, $\Delta N_{\rm eff} \lesssim 0.1$ \cite{Planck:2015xua}.  
Assuming adiabatic expansion from the QCD epoch until recombination, we have 
\begin{align}\label{eq:DNeff_analytic}
	\Delta N_{\rm eff} = \frac{\rho_a(t_\QCD)}{2 \frac{7}{8} \frac{\pi^2}{30} \left( \frac{4}{11} \right)^{4/3} T_{\rm rec}^4} \left( \frac{g_{\ast S}(t_{\rm rec}) T_{\rm rec}^3}{g_{\ast S}(t_\QCD) T_\QCD^3} \right)^{4/3} 
	\simeq \bigl( 5 \times 10^2 \GeV^{-4} \bigr) \rho_a(t_\QCD)
\end{align}
where $g_{\ast S}(t_\QCD) \simeq 20$, $T_\QCD \simeq 0.2 \GeV$, $g_{\ast S}(t_{\rm rec}) \simeq 3.91$, and $T_{\rm rec} \simeq 0.3 \eV$.  

\subsubsection{Axion relics from cosmic strings}
\label{sec:axion_from_string}

\eref{eq:rhoaQCD_string} gives the energy density of axions emitted by the network of $\pi$-strings from the PQ phase transition until the QCD epoch.  
As we have discussed in \sref{sec:strings}, the axions emitted closer to the PQ epoch remain relativistic at late times, whereas the axions emitted later become nonrelativistic soon after the QCD epoch.  
Both population are roughly equally abundant, which is the source of the logarithmic enhancement in \eref{eq:rhoaQCD_string}, and therefore we use $\rho_a(t_\QCD) \sim \mu H_\QCD^2 \, \log t_\QCD / t_\PQ$ in \erefs{eq:Oah2_analytic}{eq:DNeff_analytic} to estimate 
\begin{align}
	\Omega_a h^2 & \simeq \bigl( 1 \times 10^{-20} \bigr) \left( \frac{\rho_a(t_\QCD)}{\mu H_\QCD^2 \, \log t_\QCD / t_\PQ} \right) \left( \frac{\mu}{\pi f_\PQ^2} \right) \left( \frac{f_\PQ}{10 \TeV} \right)^2 \left( \frac{\log t_\QCD / t_\PQ}{20} \right) \label{eq:Oah2_string} \\
	\Delta N_{\rm eff} & \simeq \bigl( 2 \times 10^{-27} \bigr) \left( \frac{\rho_a(t_\QCD)}{\mu H_\QCD^2 \, \log t_\QCD / t_\PQ} \right) \left( \frac{\mu}{\pi f_\PQ^2} \right) \left( \frac{f_\PQ}{10 \TeV} \right)^2 \left( \frac{\log t_\QCD / t_\PQ}{20} \right) \label{eq:DNeff_string}
	\per
\end{align}
As we discussed in \sref{sec:strings_emission}, the tension of the $\pi$-strings is smaller than the usual axion string tension by the clockworking factor, $(f_\PQ/f_a)^2 \sim q^{-2N} \ll 1$.  
Since the tension is smaller, there is less energy available for axion emission, and the relic abundance from strings is suppressed.  
From these formulas we infer that the clockwork axion model predicts too much dark matter ($\Omega_a h^2 > \Omega_\DM h^2 \simeq 0.12$) for $f_\PQ > 9 \times 10^{13} \GeV$, and it predicts too much dark radiation ($\Delta N_{\rm eff} > 0.1$) for $f_\PQ > 2 \times 10^{17} \GeV$.  
However, the calculation breaks down for $f_\PQ > f_a$, since there is no clockworking, and we will see that $f_a < 10^{11} \GeV$ is required to avoid overproducing axion dark matter from misalignment in \sref{sec:misalignment}.  

\subsubsection{Axion dark radiation from domain walls}
\label{sec:axion_from_wall}

\begin{figure}[t]
\begin{center}
\includegraphics[width=0.50\textwidth]{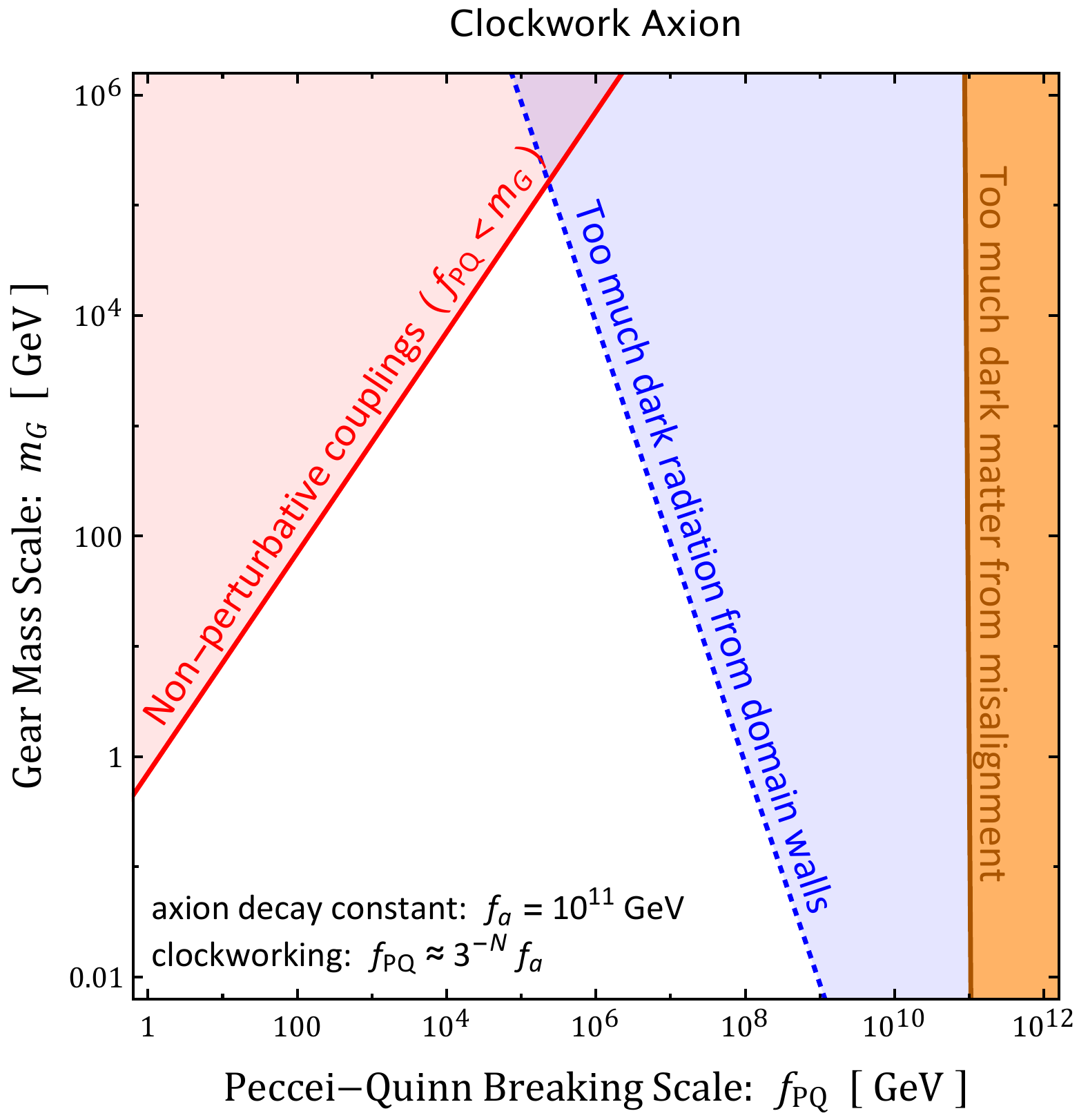} 
\caption{\label{fig:relic_abund}
Parameter space for the clockwork axion model.  The red shaded region corresponds to a non-perturbatively large value of the dimensionless coupling, $\epsilon = 2 m_G^2 / f_\PQ^2$.  In the orange shaded region there is too much axion dark matter ($\Omega_a h^2 > \Omega_\DM h^2 \simeq 0.12$) produced from misalignment with $f_\PQ = f_a$.  In this work we have calculated the axion emission from oscillating gear-field domain walls \pref{eq:DNeff_wall}, and in the blue region we find that this produces too much dark radiation $(\Delta N_{\rm eff} > 0.1)$.  Uncertainties in this calculation are large and difficult to quantify.  Assuming efficient axion emission from small-scale oscillating structure on the gear-field domain walls ($v_\omega \sim 1$), everything above the blue-dotted line is excluded.  
}
\end{center}
\end{figure}

\eref{eq:rhoaQCD_wall} gives the energy density of axions emitted by the network of $G$-walls from the PQ phase transition until the QCD epoch.  
We have discussed in \sref{sec:walls} that domain wall oscillations with $\omega \sim H_\QCD$ produce nonrelativistic axion dark matter that contributes to $\Omega_a h^2$ whereas oscillations with $\omega \sim \omega_\ast(t_\QCD) \gg H_\QCD$ produce relativistic axion dark radiation that contributes to $\Delta N_{\rm eff}$.  
Using \erefs{eq:rhoaQCD_wall}{eq:epsilon_prime} in \erefs{eq:Oah2_analytic}{eq:DNeff_analytic} we estimate 
\begin{align}
	\Omega_a h^2 & \simeq \bigl( 5 \times 10^{-26} \bigr) \left( \frac{\rho_a(t_\QCD)}{\frac{\pi^6}{N^3} v_\omega^4 \frac{m_G^4}{m_\rho^4} \frac{H_\QCD}{m_G} \sigma H_\QCD} \right) \left( \frac{v_\omega}{1} \right)^4 \left( \frac{N}{15} \right)^{-3} \nn
	& \qquad \times \left( \frac{\sigma}{8 m_G f_\PQ^2} \right) \left( \frac{m_G}{100 \GeV} \right)^3 \left( \frac{m_\rho}{1 \TeV} \right)^{-4} \left( \frac{f_\PQ}{10 \TeV} \right)^2 \label{eq:Oah2_wall} \\ 
	\Delta N_{\rm eff} & \simeq \bigl( 9 \times 10^{-8} \bigr) \left( \frac{\rho_a(t_\QCD)}{0.1 v_\omega^2 \sigma H_\QCD} \right) \left( \frac{v_\omega}{1} \right)^2 \left( \frac{\sigma}{8 m_G f_\PQ^2} \right) \left( \frac{m_G}{100 \GeV} \right) \left( \frac{f_\PQ}{10 \TeV} \right)^2 \label{eq:DNeff_wall}
	\per
\end{align}
As we anticipated in \sref{sec:walls}, the axion dark matter relic abundance arising from oscillation of Hubble-scale gear-field domain walls is very suppressed as compared with the usual QCD axion cosmology, and we will see that it is negligible when compared with the dark matter produced through misalignment.  
On the other hand, the predicted dark radiation in \eref{eq:DNeff_wall} becomes comparable to the observational upper limit, $\Delta N_{\rm eff} < 0.1$, for sufficiently large $m_G$ and $f_\PQ$.  
This translates into a constraint on the $(m_G, f_\PQ)$ parameter space appearing in \fref{fig:relic_abund}.  

Recall that $v_\omega \leq 1$ parametrizes the spectrum of perturbations on the domain wall on small scales.  
Due to the complicated structure and dynamics of the string-wall network, which we have discussed in \sref{sec:walls}, it is challenging to calculate the axion emission precisely.  
The axion relic abundance may be suppressed compared to these estimates if more energy is lost in other ways, for instance by the emission of gears.  

\subsubsection{Axion dark matter from string-wall collapse}
\label{sec:axion_from_collapse}

\eref{eq:rhoaQCD_collapse} gives the energy density of axions emitted during the collapse of the string-wall network at the QCD epoch.  
Since this axions are predominantly produced from Hubble-scale defects, they are produced with very low momentum and contribute to the dark matter.  
The predicted dark matter relic abundance is given by 
\begin{align}
	\Omega_a h^2 & \simeq \bigl( 1 \times 10^{-45} \bigr) \left( \frac{\rho_a(t_\QCD)}{N m_a^4 f_a^2 m_G^{-3} H_\QCD} \right) \left( \frac{N}{15} \right) \left( \frac{m_a}{10^{-4} \eV} \right)^{4} \left( \frac{f_a}{10^{11} \GeV} \right)^2 \left( \frac{m_G}{100 \GeV} \right)^{-3} 
	\per
\end{align}
Here we see that the axion production during the collapse of the string-wall network is entirely negligible, whereas for the usual QCD axion model this is one of the dominant contributions to the axion dark matter relic abundance.  

\subsubsection{Axion dark matter from misalignment}
\label{sec:misalignment}

Since the axion mass is negligible at the PQ phase transition, the axion field is randomly and uniformly distributed across the vacuum manifold, $a \in [0, 2\pi f_a)$ or $\theta_a = a/f_a \in [0, 2\pi)$.  
(Here we assume $N_{DW} = 1$ so that $f_g = f_a$.)  
In general the local value of $\theta_a$ at some point in space will be misaligned with the value of $\theta_a$ that minimizes the axion potential, which arises at the QCD phase transition.  
This misalignment \cite{Preskill:1982cy, Abbott:1982af, Dine:1982ah} corresponds to a local potential energy density of approximately $\rho_a = m_a(t)^2 f_a^2 (1 - \cos \theta_a )$ where $m_a(t)$ is the effective axion mass at time $t$, which grows rapidly during the QCD phase transition.  
Subsequently the axion field begins to oscillate and behave like pressureless dust.  
The corresponding relic abundance of axion dark matter is given by \cite{Fox:2004kb}
\begin{align}\label{eq:Omega_misalign}
	\Omega_a h^2 \simeq 0.2 \left( \frac{f_a}{10^{11} \GeV} \right)^{7/6} \left( \frac{\langle \theta_a^2 \rangle}{\pi^2/3} \right)
\end{align}
where the value of $\theta_a$ is averaged.  
Note that it is $f_a$ and not $f_\PQ$ that controls the misalignment contribution to the axion dark matter relic abundance.  
To avoid producing too much dark matter ($\Omega_a h^2 > \Omega_\DM h^2 \simeq 0.12$), it is necessary that $f_a \lesssim 1 \times 10^{11} \GeV$, which translates into a constraint on the parameter space appearing in \fref{fig:relic_abund}.  

\section{Conclusion}
\label{sec:Conclusion}

\begin{figure}[t]
\begin{center}
\includegraphics[width=0.49\textwidth]{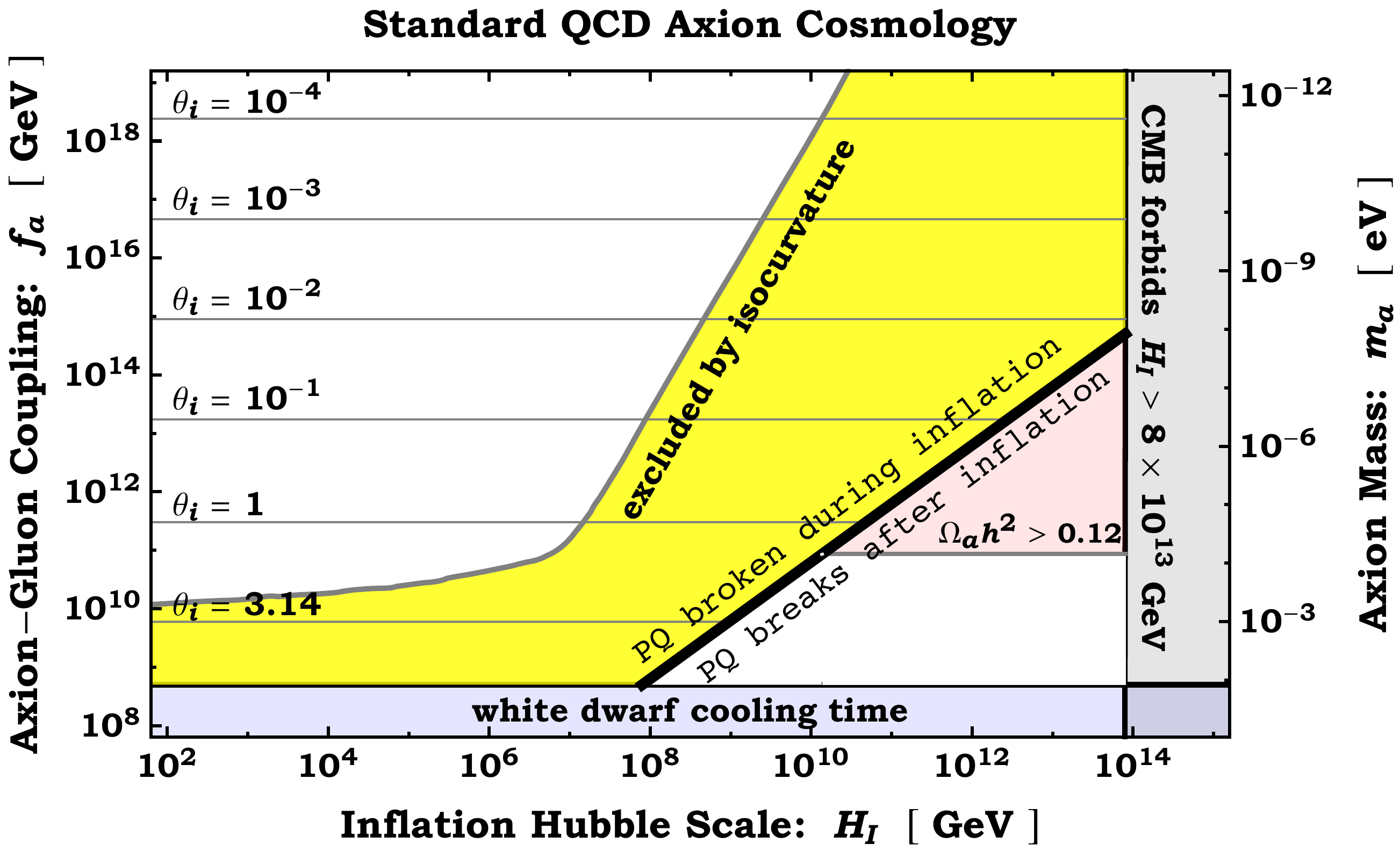} \hfill
\includegraphics[width=0.49\textwidth]{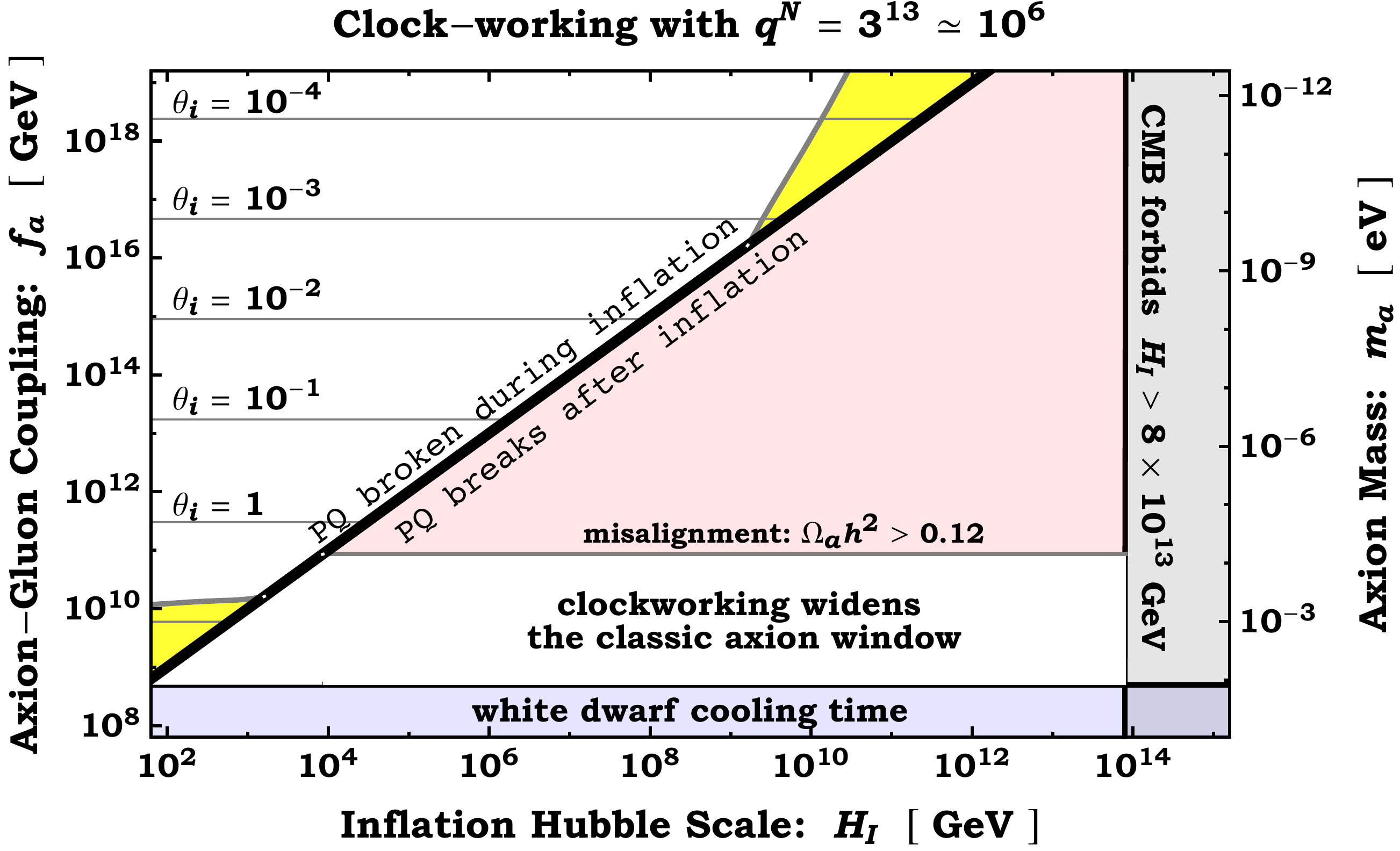} 
\caption{\label{fig:param_space}
The canonical parameter space of QCD axion cosmology (left), and how this parameter space is modified due to the clockworking (right).  
In the right panel we fix the clockworking factor, $q^N$, and vary $f_\PQ \sim f_a / q^N$.  
The diagonal boundary corresponds to $f_\PQ \sim H_I / 2\pi$.  
Clockworking widens the classic axion window, $f_a \in (10^9, 10^{11}) \GeV$, allowing lower $H_I$.  
The various shaded regions are excluded:  the yellow region is excluded by CMB limits on isocurvature \cite{diCortona:2015ldu}; the blue region is excluded by limits on white dwarf cooling time \cite{Kuster:2008zz} (observations constrain the axion-photon and axion-electron couplings, and the mapping to axion-gluon coupling is model-dependent); the gray region is excluded by CMB limits on inflationary gravitational waves $(r < 0.1)$ \cite{Planck:2015xua}; and the red region is excluded by measurements of the dark matter relic abundance (assuming the axion is cosmologically stable).  
}
\end{center}
\end{figure}

In this work we have explored some of the cosmological implications of the clockwork axion with a focus on the network of topological defects and their contribution to the axion relic abundance.  
The primary cosmological consequence of clockworking, which lowers $f_\PQ$ compared to $f_a$, is to make it easier for the PQ symmetry to be restored when the universe reheats after inflation.  
As shown in \fref{fig:param_space}, the clockworking widens the ``classic axion window'' in which the PQ symmetry becomes broken during a cosmological phase transition, and a network of topological defects is formed.  
This behavior is well known from the usual QCD axion cosmology where a network of cosmic strings forms at the time of the PQ phase transition, and the strings become bounded by axion-field domain walls at the QCD epoch.  
In the clockwork axion model, on the other hand, the PQ phase transition gives rise to both $O(N)$ ``flavors'' of cosmic strings as well as a network of domain walls composed of the new pseudoscalar gear fields; see the discussion in \sref{sec:formation}.  

The string-wall defect network emits axions as it evolves from the PQ phase transition until it eventually collapses at the QCD epoch.  
Given the rich structure of the network, it is challenging to make robust predictions for its dynamics.  
Assuming that the defect network reaches the so-called scaling regime, we calculate the relic abundance of relativistic axions as dark radiation and nonrelativistic axions as dark matter.  
Our main results are given by the formulas in \sref{sec:relics} as well as \fref{fig:relic_abund}, and we summarize the key points here.  
\begin{itemize}
	\item  Clockworking lowers the tension of the cosmic strings from $\mu \sim f_a^2$ to $\mu \sim f_\PQ^2$.  Consequently the cosmic strings carry less energy, and the axion relic abundance arising from string emission is suppressed by $(f_\PQ/f_a)^2 \sim q^{-2N}$.  (This suppression is partially offset by a competing enhancement factor, because strings loops that emit less efficiently will live longer, but the net effect still leaves a negligible axion relic abundance.)
	\item  In the clockwork axion model, domain walls composed of the gear fields form already at the time of PQ symmetry breaking.  The surface tension of these walls is $\sigma \sim m_G f_\PQ^2$, which can be comparable to the tension of axion-field walls in the usual axion cosmology, $\sigma \sim m_a f_a^2$, because $m_G \gg m_a$ even though $f_\PQ \ll f_a$.  This comparison suggests that the relic abundance of axion dark matter can be comparable to the usual estimates \pref{eq:Oah2_naive}.  However, the axion interacts only very weakly with the domain walls through the dimension-8 operator, $(\partial a) (\partial G)^3 / f_\PQ^2 m_\rho^2$.  Consequently the relic abundance of axion dark matter arising from {\it Hubble-scale} domain walls at the QCD epoch, given by \eref{eq:Oah2_wall}, is suppressed by $(f_\PQ/f_a)^2 (m_G / m_\rho)^4 (H_\QCD / m_a)$ compared to the usual axion cosmology, which makes it totally negligible.  However, the presence of {\it small-scale} oscillating features on the domain walls produces a population of relativistic axions that contribute to dark radiation; see \eref{eq:DNeff_wall} and the blue region in \fref{fig:relic_abund}.  This calculation carries a large uncertainty associated with the spectrum of oscillations on the domain wall, which would be interesting to measure with a lattice simulation of the defect network evolution.  
	\item  The defect network collapses at the QCD epoch when the axion mass arises from instanton effects.  At this time the axion potential induces a mixing between the axion and the gear fields, which provides a new avenue for axion emission.  However, the mixing is related to the ratio of the axion mass and the gear mass, and the resultant axion relic abundance is suppressed by $(m_a / m_G)^3$, which makes it negligibly small.  Most of the energy from the defect network collapse goes into emission of gear-field particles, which decay to Standard Model radiation.  
\end{itemize}

In this article we have focused on the ``simplest'' example of a clockwork axion model \cite{Kaplan:2015fuy} that is reviewed in \sref{sec:Clockwork}.  
Our results are readily generalized to other aligned axion models with a large hierarchy between $f_\PQ$ and $f_a$.  
Our key assumptions have been that (1) $f_\PQ$ is low enough that the PQ symmetry is restored during reheating after inflation and subsequently broken during a cosmological phase transition, (2) the complex structure of the defect network prohibits the formation of strings formed from the axion field, (3) the axion interacts weakly with the (gear) fields that form the domain walls, and (4) these interactions can be described by low energy effective field theory to calculate axion emission from the defects.  
More generally, in \sref{sec:comparision} we discussed various UV extensions of the clockwork axion following \rref{Agrawal:2017cmd}.  
In general there can be a delay between the formation of the string network at the PQ phase transition and the later formation of the domain walls.  
Our results for the axion relic abundance are insensitive to the UV physics, because most of the axion production is occurring at late times, at the QCD epoch.  
	
\subsubsection*{Acknowledgements}
The author is indebted to Lian-Tao Wang for introducing him to the problem studied here.  
This work was made possible through extensive and invaluable discussions with Lian-Tao Wang and Andrea Tesi.  
A.J.L. is supported at the University of Chicago by the Kavli Institute for Cosmological Physics through grant NSF PHY-1125897 and an endowment from the Kavli Foundation and its founder Fred Kavli.  

\bibliographystyle{JHEP}
\small
\bibliography{refs--clockwork}

\end{document}